%% file: black_string.tex
\documentclass[a4paper,11pt]{article}
\usepackage{marvosym}
\usepackage{comment}
\usepackage{subfigure}
\usepackage{mathrsfs} 
\usepackage[framemethod=default]{mdframed}

\newmdenv[skipabove=1pt,
skipbelow=1pt,
rightline=false,
leftline=false,
topline=false,
bottomline=false,
backgroundcolor=gray!10,
linecolor=gray,
innerleftmargin=5pt,
innerrightmargin=5pt,
innertopmargin=-10pt,
innerbottommargin=5pt,
leftmargin=0cm,
rightmargin=0cm,
linewidth=4pt]{eBox}

\catcode`\,12

\pdfoutput=1 

\usepackage{jheppub} 
\usepackage[T1]{fontenc} 

\title{From Black Strings to Fundamental Strings: Non-uniformity and Phase Transitions}

\author{Jinwei Chu}

\affiliation{Department of Physics, University of Chicago, Chicago, IL 60637, USA}

\emailAdd{jinweichu@uchicago.edu}

\abstract{
We discuss the transition between black strings and fundamental strings in the presence of a compact dimension, $\mathbb{S}^1_z$. In particular, we study the Horowitz-Polchinski effective field theory in $\mathbb{R}^d\times\mathbb{S}^1_z$, with a reduction on the Euclidean time circle $\mathbb{S}_\tau^1$. The classical solution of this theory describes a bound state of self-gravitating strings, known as a ``string star'', in Lorentzian spacetime. By analyzing non-uniform perturbations to the uniform solution, we identify the critical mass at which the string star becomes unstable towards non-uniformity along the spatial circle (i.e., Gregory-Laflamme instability) and determine the order of the associated phase transition. For $3\le d<4$, we argue that at the critical mass, the uniform string star can transition into a localized black hole. More generally, we describe the sequence of transitions from a large uniform black string as its mass decreases, depending on the value of $d$. Additionally, using the $SL(2)_k/U(1)$ model in string theory, we show that for sufficiently large $d$, the uniform black string is stable against non-uniformity before transitioning into fundamental strings. We also present a novel solution that exhibits double winding symmetry breaking in the asymptotically $\mathbb{R}^d\times\mathbb{S}^1_\tau\times\mathbb{S}^1_z$ Euclidean spacetime.
}

\input{sonerMacros}

\begin{document} 
\maketitle
\flushbottom

\section{Introduction}
The Schwarzschild black hole is a spherically symmetric solution to Einstein's theory of gravity, asymptoting to $\mathbb{R}^{d,1}$ and possessing an event horizon with $\mathbb{S}^{d-1}$ topology. Near the horizon, quantum fluctuations give rise to Hawking radiation. This quantum nature makes the black hole a promising avenue for understanding quantum gravity. As the black hole gradually evaporates through the Hawking radiation, its mass decreases over time. A natural question then arises: what does the black hole evolve into as it loses its mass during the evaporation?

If we embed this problem in string theory, an intriguing answer emerges: as the mass decreases, the black hole transitions into highly excited fundamental strings, in a regime where the horizon radius approaches the string scale~\cite{Bowick:1985af,Susskind:1993ws,Horowitz:1996nw,Horowitz:1997jc} (see also \cite{Sen:1995in,Damour:1999aw,Khuri:1999ez,Kutasov:2005rr,Giveon:2005jv,Giveon:2006pr} for relevant works). Notably, this is also the scale at which string corrections become significant, leading to the breakdown of the gravity theory. At sufficiently large mass, a collection of highly excited free strings tends to form a single long string. However, the self-gravitation of the long string may not always be negligible. The bound state formed by self-gravitating strings is commonly referred to as a ``string star''. One can study the string star in Euclidean signature with the Horowitz-Polchinski effective field theory~\cite{Horowitz:1997jc}. Altogether, these elements suggest a transition picture where the black hole first evolves into a string star and, as the mass continues to decrease, eventually becomes free strings. For recent developments, see~\cite{Brustein:2021cza,Chen:2021emg,Chen:2021dsw,Urbach:2022xzw,Balthazar:2022szl,Balthazar:2022hno,Bedroya:2022twb,Urbach:2023npi,Ceplak:2023afb,Halder:2023nlp,Agia:2023skp,Bedroya:2024uva,Santos:2024ycg}.

More generally, in addition to the noncompact spacetime $\mathbb{R}^{d,1}$, there might be compact dimensions, as is common in string theory (as well as other models such as Kaluza-Klein theory). This leads to the natural problem of understanding the transition between black holes and fundamental strings in the presence of extra compact dimensions. In this paper, we focus on the simplest case: adding one extra spatial circle to the $d+1$ noncompact dimensions. 

In the full $(d+2)$-dimensional spacetime, the Schwarzschild black hole is uplifted to a uniform black string wrapping around the spatial circle, with a horizon topology of $\mathbb{S}^{d-1}\times \mathbb{S}^1$. Gregory and Laflamme famously demonstrated that as the mass of the uniform black string decreases, it becomes unstable at a critical point $M_\text{BS,GL}$, transitioning into a non-uniform solution~\cite{Gregory:1993vy,Gregory:1994bj} (see also~\cite{Kol:2004ww} for a nice review). This instability occurs when the horizon radius approaches the size of the compact dimension, up to a $d$-dependent numerical factor. In this paper, we focus on the regime where the compact dimension is much larger than the string scale. This ensures that at the critical point of the Gregory-Laflamme instability, the gravity theory remains valid (outside the horizon).

The non-uniform solution that arises from the uniform black string via the Gregory-Laflamme instability is either a black string that is not uniform along the compact dimension, or a black hole localized in the compact dimension. As a reminder, we refer exclusively to solutions with horizon topologies of $\mathbb{S}^d$ as black holes, while the term black string refers to solutions with horizon topologies of $\mathbb{S}^{d-1}\times \mathbb{S}^1$. In general, no analytic solutions for these non-uniform black objects are known. However, when the mass is significantly below the critical point $M_\text{BS,GL}$—such that the localized black hole is much smaller than the compact dimension—it can be well-approximated by a higher-dimensional Schwarzschild black hole. 

Since the spatial circle is much larger than the localized black hole, it can effectively be treated as decompactified. Then, as the mass continues to decrease, one might speculate that the transitions proceed as they would in a noncompact $(d+2)$-dimensional spacetime. However, this speculation hinges on the assumption that the string star remains localized and fits well within the compact dimension. This assumption is problematic, as the string star can become considerably large~\cite{Chen:2021dsw}. Therefore, even after the Gregory-Laflamme instability occurs to the uniform black string, the compact nature of the extra dimension cannot be ignored. In particular, the uniform string star is also likely to play a role in the transition scenario. To fully understand this scenario, a more detailed analysis of the non-uniformity of the string star is necessary.

In this paper, we investigate the transitions that occur as the mass of a large uniform black string decreases. The paper is organized as follows: In section \ref{secreview}, we review the Schwarzschild black hole, free strings and self-gravitating strings (string star) in a $(d+1)$-dimensional noncompact spacetime. Treating $d$ as a continuous parameter, we show that as the black hole mass decreases, for $2<d<4$, it first transitions into a string star, which then gradually evolves into free strings, while for $d>4$, the black hole transitions directly into free strings. In section \ref{secGL}, we introduce an additional spatial circle and review the Gregory-Laflamme instability of the uniform black string. 

In section \ref{secnHP}, we study the string star in $\mathbb{R}^{d,1}\times\mathbb{S}^1$ by Wick rotating to Euclidean spacetime, reducing on the Euclidean time circle, and analyzing the Horowitz-Polchinski effective field theory in $\mathbb{R}^d\times\mathbb{S}^1$. Taking the uniform solution to be the background and analyzing non-uniform solutions perturbatively to the next-to-leading order, we determine the critical point of the Gregory-Laflamme instability for the string star and the order of the phase transition. We highlight that for $2<d<4$, the uniform string star is stable for masses smaller than the critical mass $M_\text{HP,GL}$. This suggests that a non-uniform solution can transition into the uniform string star as the mass decreases, in contrast to the behavior observed in the black string case. We also discuss possible non-uniform solutions resulting from the Gregory-Laflamme instability of the uniform string star in this section. Interestingly, for $3< d<4$, we find that at the critical point $M_\text{HP,GL}$, the uniform string star transitions directly into a localized black hole, rather than into a non-uniform string star. 

\begin{figure}
	\centering
\includegraphics[scale=0.8]{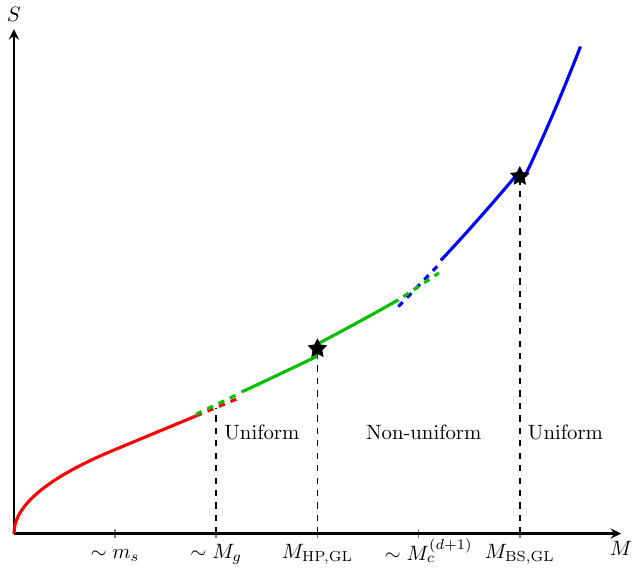}
\caption{\label{entropiesd3}Entropies of black objects (blue lines), string stars (green lines) and free strings (red line) in $\mathbb{R}^{d,1}\times\mathbb{S}^1_z$ for $2<d<3$. The stars mark the critical points of the Gregory-Laflamme instabilities. See section \ref{seccomplete} for more details.}
\end{figure}
In section \ref{seccomplete}, we describe the resulting qualitative pictures, which vary depending on the value of $d$ (readers primarily interested in this overview may choose to skip section \ref{secnHP} at a first reading). For instance, when $d$ is between 2 and 3, the phase structure is particularly rich. Specifically, as shown in figure \ref{entropiesd3}, the following sequence of phase transitions takes place as the mass decreases: uniform black string $\to$ localized black hole $\to$ localized string star $\to$ uniform string star $\to$ free string. In this section, we also include an analysis of the large $d$ case, which can be studied exactly in string theory using the $SL(2,R)_k/U(1)$ WZW model~\cite{Emparan:2013xia,Chen:2021emg,Halder:2024gwe}. We find that for sufficiently large $d$, the Gregory-Laflamme instability does not occur before the uniform black string transitions into highly excited fundamental strings. We conclude the paper and discuss future work in section \ref{secdis}.

Additionally, in appendix \ref{sec2HP}, we analyze the Horowitz-Polchinski effective field theory on the thermal Kaluza-Klein bubble, which yields an interesting string star saddle that breaks the winding symmetries along both the Euclidean time circle and the extra spatial circle.
\section{Review: transition in $d+1$ noncompact dimensions}\label{secreview}
In string theory, a transition between black holes and fundamental strings can occur when the size of the black hole becomes comparable to the string length~\cite{Bowick:1985af,Susskind:1993ws,Horowitz:1996nw,Horowitz:1997jc}. We review this transition in detail in this section, assuming that the spacetime has $d+1$ noncompact dimensions.
\subsection{Schwarzschild black hole}\label{secBH}
Let us first consider the $(d+1)$-dimensional Schwarzschild black hole, which is a solution to Einstein's theory of gravity that has a spacetime background with zero scalar curvature. The metric is
\begin{equation}
\label{schw}
    ds^2=-\left(1-\left(\frac{r_H}{r}\right)^{d-2}\right)dt^2+\frac{dr^2}{1-\left(\frac{r_H}{r}\right)^{d-2}}+r^2d\Omega_{d-1}^2\ ,
\end{equation}
where $r_H$ denotes the horizon radius and $d\Omega_{d-1}^2$ represents the line element of the unit $(d-1)$-sphere. The Schwarzschild solution exists only for $d\ge 3$. However, in the following we will treat $d$ as a continuous parameter, in which case (\ref{schw}) exists for $d>2$.

For large $r$, the metric (\ref{schw}) approaches flat spacetime. The curvature\footnote{Note that the scalar curvature vanishes identically for the Schwarzschild metric (\ref{schw}). However, it does not mean that the spacetime is flat. For example, the Kretschmann scalar $K\equiv R^{\mu\nu\rho\sigma}R_{\mu\nu\rho\sigma}$ is nonzero in this case.} increases as $r$ decreases and diverges at the singularity $r=0$, where classical gravity breaks down. To suppress quantum gravity effects near the horizon, the horizon radius $r_H$ must be much larger than the Planck length $l_p$. Furthermore, string theory introduces another length scale, the string length $l_s\equiv \sqrt{\alpha'}$, at which string corrections become important. This length scale is related to $l_p$ via $l_p=g_s^{2/(d-1)}l_s$, where $g_s$ denotes the string coupling. In a weakly-coupled string theory ($g_s\ll 1$), $l_p\ll l_s$. Thus, as long as $r_H\gg l_s$, the gravity result outside the horizon is reliable.

The ADM mass of the black hole can be extracted from the fall-off coefficient of $g_{tt}$ in (\ref{schw}). Specifically, it is given by
\begin{equation}
\label{M}
    M=\frac{(d-1)\omega_{d-1}}{16\pi G_N^{(d+1)}}r_H^{d-2}\ .
\end{equation}
Here $\omega_{d-1}$ denotes the area of the unit $(d-1)$-sphere. The $(d+1)$-dimensional Newton's constant, $G_N^{(d+1)}$, is related to the string length and string coupling via 
\begin{equation}
\label{GNls}
   G_N^{(d+1)}\equiv l_p^{d-1}=g_s^2l_s^{d-1} 
\end{equation}
up to a numerical prefactor of order one. 

As mentioned earlier, the metric remains valid outside the horizon when $r_H \gg l_s$. Expressed in terms of the mass $M$, (\ref{M}), this condition corresponds to the regime
\begin{equation}
\label{Mc}
    M\gg M_c\equiv \frac{l_s^{d-2}}{G_N^{(d+1)}}=\frac{1}{g_s^2l_s}\ .
\end{equation}
Here, $M_c$ represents the mass scale of the black hole when its horizon radius is of order $l_s$. As $M$ approaches $M_c$, the metric receives significant string corrections, and the fall-off coefficient of $g_{tt}$, in terms of the horizon radius $r_H$, is no longer $r_H^{d-2}$~\cite{CALLAN1989673}. Here, the radial coordinate $r$ is defined as the radius of the $(d-1)$-sphere, i.e., $g_{\Omega\Omega}=r^2$. Consequently, the relation between $M$ and $r_H$, as expressed in (\ref{M}), becomes unreliable.

It is well known that a black hole has an entropy which is proportional to the area of the horizon (if there are no string corrections). For the Schwarzschild black hole (\ref{schw}), the entropy is given by
\begin{equation}
\label{S}
    S=\frac{\omega_{d-1}r_H^{d-1}}{4G_N^{(d+1)}}\ .
\end{equation}
Combining (\ref{M}) and (\ref{S}), we obtain a formula for the black hole entropy in terms of its mass,
\begin{equation}
\label{SBH}
    S=\frac{1}{4}\left(\frac{G_N^{(d+1)}}{\omega_{d-1}}\right)^{\frac{1}{d-2}}\left(\frac{16\pi M}{d-1}\right)^{\frac{d-1}{d-2}}
\end{equation}
with the regime of validity (\ref{Mc}). As we will review in the next subsection, naively extrapolating (\ref{SBH}) to the regime $M\sim M_c$ yields a result that matches the entropy of free strings, which is valid in the opposite regime $M\ll M_c$. This suggests that the black hole undergoes a transition to a free string state near the mass $M_c$.

\subsection{Free string}\label{secfree}
A collection of free strings tends to form a single long string when its mass is much larger than the string mass, i.e., $M\gg m_s\equiv 1/l_s$. The entropy of the long string exhibits Hagedorn behavior. In the leading order of $Ml_s$, it is given by
\begin{equation}
\label{Sfree}
    S=\beta_HM\ .
\end{equation}
Here $\beta_H$ is known as the (inverse) Hagedorn temperature. At this critical temperature, the partition function of the string canonical ensemble,
\begin{equation}
    Z=\int_0^\infty dM e^{-\beta M+S}\sim\int_0^\infty dM e^{(\beta_H-\beta) M}\ ,
\end{equation}
diverges, which indicates the breakdown of the canonical ensemble. The exact value of $\beta_H$ depends on the type of the string theory. For instance,
\begin{equation}
\label{betaH}
    \beta_H^{\text{bosonic}}=4\pi l_s\ ,\quad \beta_H^{\text{type II}}=2\sqrt{2}\pi l_s\ .
\end{equation}

It is observed that to calculate the free energy of the string canonical ensemble, we can Wick rotate $\tau=it$ and consider a Euclidean field theory for string propagation on $\mathbb{R}^d\times\mathbb{S}^1_\tau$~\cite{Polchinski:1985zf}. Here, $\mathbb{S}^1_\tau$ represents the Euclidean time circle, with its period identified with the inverse temperature $\beta$. Consider a string tachyon wound once around the $\tau$ circle. This tachyon is not projected out, even in type II string theory. This is because fermions obey anti-periodic boundary conditions along the $\tau$ circle, and for odd winding numbers, the GSO projection is opposite to the usual one~\cite{Atick:1988si}. 

The mass of the winding one tachyon is given by
\begin{equation}
\label{m2}
    m^2=\frac{\beta^2-\beta^2_H}{(2\pi\alpha')^2}\ ,
\end{equation}
where $\beta_H$ is precisely the Hagedorn temperature specified in (\ref{betaH}). Thus, the Hagedorn temperature can be interpreted as the critical Euclidean time period at which the winding one tachyon becomes massless.
\begin{figure}
	\centering
\includegraphics[scale=0.7]{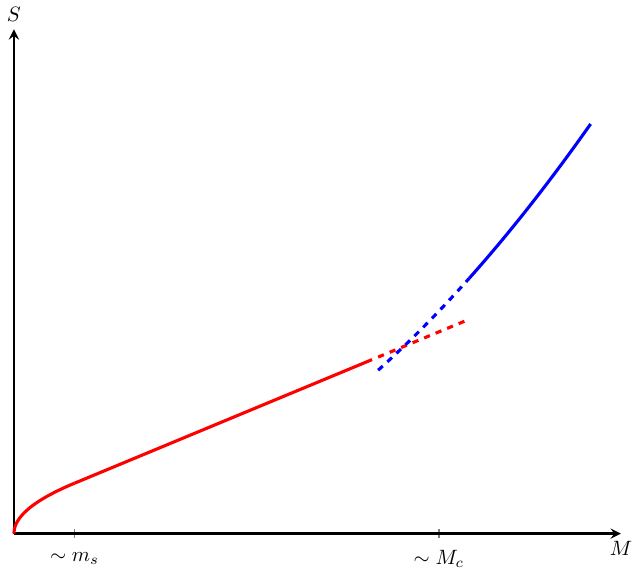}
\caption{\label{entropies1}The entropy of the black hole (in blue), given by (\ref{SBH}), and that of the free string (in red), given by (\ref{Sfree}). The dashed lines represent naive extrapolations of the formulas beyond their regimes of validity. $M_c$ is far to the right of $m_s$ on the horizontal axis, as is clear from the definition (\ref{Mc}) with $g_s\ll 1$.}
\end{figure}

We now relate the free string to the Schwarzschild black hole discussed in the previous subsection. Let us compare the entropy of the free string (\ref{Sfree}) with that of the Schwarzschild black hole (\ref{SBH}), as illustrated in figure \ref{entropies1}. Note that the free string result (\ref{Sfree}) is derived for $g_s=0$. For small but nonzero $g_s$, we assume that (\ref{Sfree}) is still applicable over a sufficiently large range. Besides, the black hole result is unreliable at $M_c$, yet its extrapolated value becomes of the same order as the extrapolation of the free string entropy (\ref{Sfree}) around that point. This observation naturally leads to the conjecture that the black hole transitions into the free string around the mass $M_c$~\cite{Bowick:1985af,Susskind:1993ws,Horowitz:1996nw,Horowitz:1997jc}.

The discussion on the transition between black holes and fundamental strings remains incomplete. Specifically, the self-gravitation of the long string has been neglected in the previous analysis. According to \cite{Horowitz:1997jc}, self-gravitation becomes significant when $g_s(Ml_s)^{\frac{6-d}{4}}\gtrsim 1$ for $M\gg m_s$. For $d<6$, this implies that the self-gravitation is negligible only if 
\begin{equation}
\label{Mg}
    M\ll M_g\equiv \frac{g_s^{\frac{4}{d-6}}}{l_s}\ .
\end{equation}
Interestingly, when $d<4$, comparing $M_g$ with $M_c$ reveals that $M_g\ll M_c$. Consequently, there is an intermediate region $M_g\lesssim M\lesssim M_c$ where the solution must be reexamined, and the string entropy formula given by (\ref{Sfree}) needs to be modified. In fact, as we will discuss in the next subsection, there exists a classical solution which takes the self-gravitation into account and is valid in the regime $M_g\ll M\ll M_c$. On the other hand, when $4<d<6$, we find $M_g\gg M_c$, indicating that self-gravitation is not expected to play a significant role in the transition from the black hole to the free string.

For $d>6$, when the string coupling is small and $M\gg m_s$, $g_s(Ml_s)^{\frac{6-d}{4}}$ never reaches order one. Therefore, the free string description remains valid along the red solid line in figure \ref{entropies1}.

\subsection{Self-gravitating string I: Horowitz-Polchinski solution}\label{sechp}
Following \cite{Horowitz:1997jc}, we can study the self-gravitating string with a $d$-dimensional effective field theory at a fixed temperature $\beta$, which describes a canonical ensemble, near the Hagedorn temperature $\beta_H$. To derive the effective field theory, we first Wick rotate the time $\tau=it$, thereby converting to the Euclidean spacetime $\mathbb{R}^d\times\mathbb{S}^1_\tau$. Here again $\mathbb{S}^1_\tau$ is the Euclidean time circle, with the period of $\beta$. We will subsequently reduce the theory on the $\tau$ circle to formulate a $d$-dimensional action for the effective field theory. It is important to stress that our primary interest lies in the microcanonical ensemble; however, we use the canonical ensemble here as it is convenient for deriving the thermodynamic quantities.

As mentioned in the previous subsection, in the Euclidean signature, there is a winding one tachyon with the mass (\ref{m2}). Approaching the Hagedorn temperature from below, this winding one tachyon becomes light. So, we need to include it in the effective field theory. 

Let us denote the field for the winding one tachyon as $\chi(x)$, which depends on the spatial coordinates $x$. This is a complex scalar field, with the conjugate $\chi(x)^*$ corresponding to the string tachyon with a winding number of $-1$. 

The Euclidean manifestation of the self-gravitation of the long string arises from the coupling between the field $\chi$ and $g_{\tau\tau}$, specifically through the modification of the mass of $\chi$ due to variations in $g_{\tau\tau}$. In particular, when $g_{\tau\tau}(x)$ varies across the spatial dimensions, the length of the time circle, given by $\beta \sqrt{g_{\tau\tau}}$, changes locally. Consequently, the mass formula (\ref{m2}) must be modified to
\begin{equation}
\label{meff}    m^2_{\text{eff}}(x)=\frac{\beta^2g_{\tau\tau}(x)-\beta^2_H}{(2\pi\alpha')^2}\ .
\end{equation}

It is now clear that the $d$-dimensional effective field theory includes the winding tachyon field $\chi$, along with the gravitational fields, specifically the $(d+1)$-dimensional graviton $g_{\mu\nu}$ and the dilaton $\phi_{d+1}$, while other heavy fields can be neglected. For convenience, we express the $\tau\tau$ component of the metric as $g_{\tau\tau}=e^{2\varphi}$. and define a $d$-dimensional dilaton as $\phi_d=\phi_{d+1}-\varphi/2$. Then, we can reduce on the $\tau$ circle and obtain the effective action
\begin{equation}
	\label{Sphichi}
	\begin{split}
 	I_d=\frac{\beta}{16\pi G_N^{(d+1)}}\int d^dx\sqrt{g}e^{-2\phi_d}\left[-\mathcal{R}-4(\nabla\phi_d)^2+(\nabla\varphi)^2+|\nabla \chi|^2+m_\text{eff}^2|\chi|^2\right]\ .
	\end{split}
	\end{equation}
Here $g$ is the determinant of the $d$-dimensional string metric and $\mathcal{R}$ is the corresponding scalar curvature. Higher-order terms, such as $|\chi|^4$ and $\varphi|\nabla\chi|^4$,\footnote{From the tree-level calculation of the three-point amplitude, the higher-order term $\varphi|\nabla\chi|^2$ can be excluded~\cite{Brustein:2021ifl}. I thank Yoav Zigdon for bringing this to my attention.} have been neglected under the assumption that the fields included in (\ref{Sphichi}) are small and vary slowly with respect to the string length.

Given the smallness of the fields, we can expand $e^\varphi=1+\varphi+\cdots$ and further neglect higher-order terms in $\varphi$. We also impose the condition that $\varphi$ vanishes asymptotically, ensuring that the asymptotic length of the $\tau$ circle is given by $\beta$. With these, the effective mass (\ref{meff}) can be expanded as
\begin{equation}
\label{meff2}
    m^2_{\text{eff}}(x)=m_\infty^2+\frac{\kappa}{\alpha'}\varphi+\cdots\ ,
\end{equation}
where $m_\infty^2$ is the mass at infinity, as given by (\ref{m2}). The linear coefficient in (\ref{meff2}) is
\begin{equation}
\label{kappa0}
    \kappa=\frac{\beta^2}{2\pi^2\alpha'}\ .
\end{equation}
At $\beta_H$, using (\ref{betaH}), we find
\begin{equation}
\label{kappa}
    \kappa^{\text{bosonic}}=8\ ,\quad \kappa^{\text{type II}}=4\ .
\end{equation}

When the inverse temperature is close to $\beta_H$, i.e. $\alpha'm_\infty^2\ll 1$, we can further neglect the $d$-dimensional graviton and the dilaton $\phi_d$ in the effective action. This is because, as can be seen from (\ref{Sphichi}) and (\ref{meff2}), the coupling between $\varphi$ and $\chi$ is of order $\alpha'^{-1}$, whereas the couplings of the graviton and $\phi_d$ to $\chi$ are of order $m_\infty^2$. So, the coupling between $\varphi$ and $\chi$ dominates, allowing us to truncate the action (\ref{Sphichi}) to retain only the leading interaction term $\varphi|\chi|^2$. Assuming a flat string metric and a vanishing dilaton $\phi_d$ in the spatial dimensions, we obtain the Horowitz-Polchinski effective action~\cite{Horowitz:1997jc}
\begin{equation}
	\label{Sphichid}
	\begin{split}
 	I_d=\frac{\beta}{16\pi G_N^{(d+1)}}\int d^dx\left[(\nabla\varphi)^2+|\nabla \chi|^2+\left(m_\infty^2+\frac{\kappa}{\alpha'}\varphi\right)|\chi|^2\right]\ ,
	\end{split}
	\end{equation}
where $\kappa$ can be approximated by its value at $\beta_H$, as given in (\ref{kappa}).

We now study the classical solution of the effective action (\ref{Sphichid}), namely the Horowitz-Polchinski solution, which describes a string star. We are interested in solutions that preserve spherical symmetry, implying that the fields depend only on the radial coordinate $r\equiv \sqrt{(x_1)^2+\cdots+(x_d)^2}$. In addition, we assume that $\chi$ is real for all $r$. Under these conditions, the equations of motion for (\ref{Sphichid}) become
        \begin{equation}
        \label{eomHP}
	\begin{split}
 	\nabla^2\chi(r)&=\chi''(r)+\frac{d-1}{r}\chi'(r)=\left(m_{\infty}^2+\frac{\kappa}{\alpha'}\varphi(r)\right)\chi(r) \ ,\\
 	\nabla^2\varphi(r) &=\varphi''(r)+\frac{d-1}{r}\varphi'(r)=\frac{\kappa}{2\alpha'}\chi^2(r)\ .
	\end{split}
	\end{equation}

The normalizable solutions for $\chi$ and $\varphi$ should satisfy boundary conditions such that both fields remain finite at $r=0$ and go to zero at large $r$. Consequently, from (\ref{eomHP}), the derivatives of the fields must be zero at $r=0$. At large $r$, the asymptotic behaviors of the fields are given by
\begin{equation}
\label{asym}
    \chi(r)\sim A_\chi \frac{e^{-m_\infty r}}{r^{\frac{d-1}{2}}}\ ,\quad \varphi(r)\sim -\frac{C_\varphi}{r^{d-2}}\ .
\end{equation}
We can use the symmetry of (\ref{eomHP}) under $\chi\to-\chi$ to set $A_\chi$ positive without loss of generality. Furthermore, integrating the second equation in (\ref{eomHP}) yields an expression for the fall-off coefficient of $\varphi$,
\begin{equation}
\label{intchi}
    C_\varphi=\frac{\kappa}{2(d-2)\alpha'}\int_0^\infty dr\ r^{d-1}\chi^2(r)\ ,
\end{equation}
which is also positive.

While an analytic solution to (\ref{eomHP}) is not known, a rescaling can be used to simplify the differential equations. By redefining the variables as
\begin{equation}
\label{resHP}
	\begin{split}
	r&=\hat{r}/m_\infty\ ,\\
 	\chi (r)&=\frac{\sqrt{2}\alpha' }{\kappa}m^2_{\infty}\hat{\chi}(\hat{r})\ ,\\
 	\varphi (r)&=\frac{\alpha'}{\kappa }m^2_{\infty}\hat{\varphi}(\hat{r})\ ,
	\end{split}
	\end{equation}
(\ref{eomHP}) becomes
\begin{equation}
\label{eomHPres}
	\begin{split}
 	\partial^2_{\hat{r}}\hat{\chi}+\frac{d-1}{\hat{r}}\partial_{\hat{r}}\hat{\chi} &=(1+\hat{\varphi})\hat{\chi} \ ,\\
 	\partial^2_{\hat{r}}\hat{\varphi}+\frac{d-1}{\hat{r}}\partial_{\hat{r}}\hat{\varphi} &=\hat{\chi}^2\ .
	\end{split}
	\end{equation}
It is clear that the rescaled equations (\ref{eomHPres}) is independent of $m_\infty$ and $\alpha'$, and the same holds for the solutions $\hat\chi$ and $\hat\varphi$. In other words, all dependence of $\chi(r)$ and $\varphi(r)$ on $m_\infty$ and $\alpha'$ is explicitly captured in the redefinitions (\ref{resHP}).

We can solve (\ref{eomHPres}) numerically (see, for example, \cite{Balthazar:2022szl}). The ``ground-state'' solution (i.e., the solution with no nodes) is uniquely determined by the boundary conditions. It turns out that such a solution exists only for $2<d<6$~\footnote{However, in some curved spacetimes, string star solutions may exist for $d> 6$, such as in the thermal Kaluza-Klein bubble (see appendix \ref{sec2HP}) and the Anti-de Sitter spacetime (see~\cite{Urbach:2022xzw})} \cite{Chen:2021dsw}. In addition, one can extend the solution slightly beyond $d=6$ by including quartic terms in the action \cite{Balthazar:2022hno}. However, we expect that only the solutions for $d<4$ would provide new insights into the transition from black holes to fundamental strings, since as discussed in the previous subsection, the self-gravitating effect is negligible during the transition for $d>4$. We will confirm that this is indeed the case for $4<d< 6^+$, where the string star solutions are known.

We now derive the entropy-mass relation for the string star in the case where $d<6$. The ADM mass can be obtained from the asymptotic form of the Einstein metric component $g_{E,\tau\tau}$. From $\phi_d\equiv \phi_{d+1}-\frac{\varphi}{2}= 0$, we express the metric component in the Einstein frame as
\begin{equation}
    g_{E,\tau\tau}=e^{-4\frac{\phi_{d+1}}{d-1}}g_{\tau\tau}=e^{2\frac{d-2}{d-1}\varphi}\ .
\end{equation}
At large $r$, utilizing the asymptotic behavior of $\varphi$ given in (\ref{asym}), we find that
\begin{equation}
   g_{E,\tau\tau}\sim 1-2\frac{d-2}{d-1}\frac{C_\varphi}{r^{d-2}}\ .
\end{equation}
From this expression, we can extract the ADM mass as
\begin{equation}
\label{MC}
    M=\frac{(d-2)\omega_{d-1}}{8\pi G_N^{(d+1)}}C_\varphi\ .
\end{equation}

It is also useful to define a rescaled fall-off coefficient $\hat C_\varphi$ through the asymptotic form of $\hat\varphi$
\begin{equation}
\label{hatC}
    \hat\varphi\sim -\frac{\hat C_\varphi}{\hat r^{d-2}}\ .
\end{equation}
Note that $\hat C_\varphi$ only depends on $d$. Then, the dependence of $M$ on $m_\infty$ can be made more explicit by combining (\ref{resHP}), (\ref{MC}) and (\ref{hatC}). This leads to
\begin{equation}
\label{MhatC}
    M=\frac{(d-2)\omega_{d-1}}{8\pi G_N^{(d+1)}}\frac{\alpha'}{\kappa}m_\infty^{4-d}\hat C_\varphi\ .
\end{equation}

The entropy can be calculated using the formula 
\begin{equation}
\label{SMF}
    S=\beta M-\beta F\ .
\end{equation}
In this expression, the free energy $F$ can be derived from the on-shell action. By performing a scaling analysis, as demonstrated in \cite{Chen:2021dsw}, we obtain
\begin{equation}
\label{FId}
\begin{split}
    F=\frac{I_d}{\beta}=&\frac{2}{6-d}\frac{m_\infty^2\omega_{d-1}}{16\pi G_N^{(d+1)}}\int_0^\infty dr\ r^{d-1}\chi^2(r)\\
    =&\frac{d-2}{6-d}\frac{\omega_{d-1}\alpha'm_\infty^2}{4\pi G_N^{(d+1)}\kappa}C_\varphi\\
    =&\frac{2}{6-d}\frac{\alpha'm_\infty^2}{\kappa}M\ ,
\end{split}
\end{equation}
where we have used (\ref{intchi}) in the second line and (\ref{MC}) in the third line.

We now apply the formula (\ref{SMF}) and expand it in the leading orders of $\beta-\beta_H$. This leads to
\begin{equation}
\label{SHP}
\begin{split}
    S=&\beta_HM+(\beta-\beta_H)M-\beta_HF+O\left((\beta-\beta_H)^2\right)\\
    =&\beta_HM+\left(\frac{2(\pi\alpha')^2}{\beta_H}-\beta_H\frac{2}{6-d}\frac{\alpha'}{\kappa}\right)m_\infty^2M+O\left((\beta-\beta_H)^2\right)\\
    =&\beta_HM+\frac{2(4-d)(\pi\alpha')^2}{(6-d)\beta_H}\left(\frac{8\pi G_N^{(d+1)}\kappa}{(d-2)\omega_{d-1}\alpha'\hat C_\varphi}\right)^{\frac{2}{4-d}}M^{\frac{6-d}{4-d}}+O\left((\beta-\beta_H)^2\right)\ ,
\end{split}
\end{equation}
where in the last step we have used (\ref{kappa0}) and expressed $m_\infty$ in terms of $M$ via (\ref{MhatC}).

The regime of validity in $M$ is constrained by two values. First, as mentioned above, the Horowitz-Polchinski effective action (\ref{Sphichid}) is valid when $\alpha'm_\infty^2\ll 1$. Second, the quantum effects are suppressed only when $I_d\gg 1$. Therefore, using (\ref{MhatC}) and (\ref{FId}), we find the regime of validity as follows:
\begin{equation}
\label{regd4}
    M_g\ll M\ll M_c\ ,\quad d<4
\end{equation}
and
\begin{equation}
\label{reg4d6}
    M_c\ll M\ll M_g\ ,\quad 4<d<6\ .
\end{equation}
Recall that the definitions of $M_c$ and $M_g$ are given in (\ref{Mc}) and (\ref{Mg}), respectively.

\begin{figure}
	\centering
	\subfigure[]{
	\begin{minipage}[t]{0.45\linewidth}
	\centering
	\includegraphics[width=2.5in]{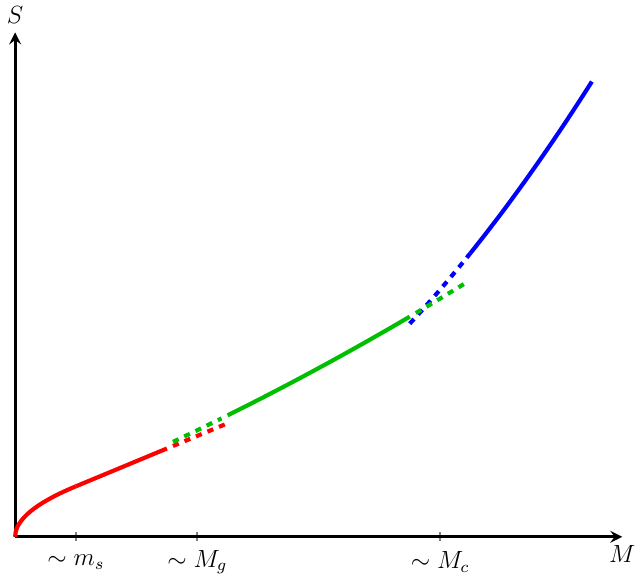}\label{entropiesdl4}
	\end{minipage}}
	\subfigure[]{
	\begin{minipage}[t]{0.45\linewidth}
	\centering
	\includegraphics[width=2.5in]{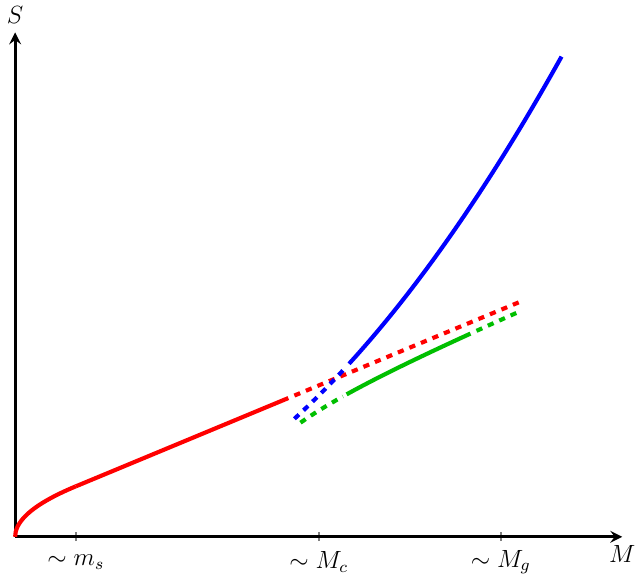}\label{entropiesdg4}
	\end{minipage}}
	\centering
\caption{\label{entropiesd4}Entropies of the Schwarzschild black hole (\ref{SBH}), the string star (\ref{SHP}) and the free string (\ref{Sfree}), represented by blue, green and red lines respectively, for (a) $d<4$ and (b) $4<d<6$. Similar figures are also sketched in~\cite{Chen:2021dsw}.}
\end{figure}

We now consider the microcanonical ensemble. For $d<4$, the regime (\ref{regd4}) contains both the string star and the free string. We can compare the Horowitz-Polchinski entropy (\ref{SHP}) with the free string entropy (\ref{Sfree}). Note that there is a next-to-leading-order correction to (\ref{Sfree}), which is proportional to $-\log M/m_s$. Since the second term in (\ref{SHP}) is positive for $d<4$, the string star has a higher entropy than the free string. This is consistent with the expectation that self-gravitation is turned on at $M\gtrsim M_g$ for $d<4$, as mentioned in the previous subsection. Therefore, in the case of $d<4$, as the mass decreases to $M\sim M_c$, the black hole transitions into a string star rather than a free string. The relevant entropy change is sketched in figure \ref{entropiesdl4}.

For $4<d<6$, in the regime of validity (\ref{reg4d6}), the Schwarzschild black hole solution is also reliable. Furthermore, the entropy expressed in (\ref{SHP}) is much lower than that of the black hole, (\ref{SBH}). Therefore, as anticipated for $d>4$, the effects of self-gravitation are negligible during the transition from the black hole to the free string (or becoming significant only in the transition regime $M\sim M_c$). For completeness, we add the entropy curve for the string star to figure \ref{entropies1}, as shown in figure \ref{entropiesdg4}.

For $d\ge 6$, we similarly expect that the self-gravitation of the string is negligible throughout the regime where the free string description remains valid. To verify this, we examine the case of $d$ close to 6 in the next subsection, where it is necessary to consider the quartic corrections to the Horowitz-Polchinski effective action~\cite{Balthazar:2022hno}.
\subsection{Self-gravitating string II: $d$ close to 6}\label{secd6}
In this subsection, we discuss the self-gravitating string for $d$ near 6, either slightly above or below. First, when $m_\infty$ and $d-6$ are exactly zero, there exists an analytic solution of (\ref{eomHP})~\cite{Balthazar:2022szl},
\begin{equation}
\label{d6HP}
    \chi(r)=-\sqrt{2}\varphi(r)=\frac{\chi(0)}{\left(1+\frac{\kappa}{24\sqrt{2}\alpha'}\chi(0)r^2\right)^2}\ .
\end{equation}
For small deviations, namely $|d-6|,\alpha'm_\infty^2\ll 1$, we can take (\ref{d6HP}) as the leading-order solution. In the following analysis, we will neglect subleading corrections to the solution, which are of higher order in $|d-6|$ and $\alpha'm_\infty^2$, for simplicity.

From (\ref{d6HP}), the fall-off coefficient of $\varphi$ can be read off as
\begin{equation}
    C_\varphi=\frac{\alpha'^2}{\kappa^2}\frac{576\sqrt{2}}{\chi(0)}\ .
\end{equation}
Using (\ref{MC}), this leads to an expression for the ADM mass,
\begin{equation}
\label{Mchi0}
    M=\frac{\omega_5}{2\pi G_N^{(7)}}C_\varphi=\frac{288\sqrt{2}\omega_5}{\pi G_N^{(7)}\chi(0)}\frac{\alpha'^2}{\kappa^2}\ ,
\end{equation}
where we take $d=6$ since we are focusing solely on the leading-order result.

$\chi(0)$ is not a free parameter; rather, it must be determined by extremizing the free energy. Substituting this solution into the effective action (\ref{Sphichid}), we find, to leading orders in $\alpha'm_\infty^2$ and $d-6$,
\begin{equation}
\label{Idmd}
\begin{split}
    F=\frac{\omega_5}{16\pi G_N^{(7)}}\bigg[ \frac{1152\alpha'^2}{5\kappa^2}+m_\infty^2\frac{4608\sqrt{2}\alpha'^3}{\chi(0)\kappa^3}+(6-d)\frac{576\alpha'^2}{5\kappa^2}\ln\frac{\chi(0)\kappa}{\alpha'}+\cdots\bigg]\ .
\end{split}
\end{equation}
Here, the ellipsis represents higher-order terms in $\alpha'm_\infty^2$ and $d-6$. We have also omitted a $\chi(0)$-independent term of order $d-6$ for conciseness.

We notice that the $\chi(0)$-dependent terms in (\ref{Idmd}) vanish as $\alpha'm_\infty^2$ and $d-6$ approach zero. This suggests that higher-order terms in the effective action cannot be neglected, as they could give non-vanishing contributions to (\ref{Idmd}) even in the limit $\alpha'm_\infty^2,d-6\to 0$. As shown in~\cite{Brustein:2021ifl}, the next-to-leading-order interactions in the effective action are given by quartic terms, including $\varphi^2|\chi|^2$ and $|\chi|^4$. More explicitly, the effective action incorporating these quartic terms is
\begin{equation}
\label{Sphichi4}
	\begin{split}
 	I_d=\frac{\beta}{16\pi G_N^{(d+1)}}\int d^dx\left[(\nabla\varphi)^2+|\nabla \chi|^2+\left(m_\infty^2+\frac{\kappa}{\alpha'}\varphi+\frac{\kappa}{\alpha'}\varphi^2\right)|\chi|^2+\frac{\kappa}{4\alpha'}|\chi|^4\right]\ .
	\end{split}
	\end{equation}
 
Next, we plug the solution (\ref{d6HP}) in the action (\ref{Sphichi4}). The resulting $\chi(0)$-dependent terms in the free energy are
\begin{equation}
\label{Fdmd}
    F=\frac{\omega_5}{16\pi G_N^{(7)}}\bigg[ m_\infty^2\frac{4608\sqrt{2}\alpha'^3}{\chi(0)\kappa^3}+(6-d)\frac{576\alpha'^2}{5\kappa^2}\ln\frac{\chi(0)\kappa}{\alpha'}+\frac{3456\sqrt{2}\alpha'^2\chi(0)}{35\kappa^2}+\cdots\bigg]\ ,
\end{equation}
where the ellipsis stands for higher-order terms in $\alpha'm_\infty^2$ and $d-6$ (we also omit $\chi(0)$-independent terms such as the first term in (\ref{Idmd})). Extremizing (\ref{Fdmd}) with respect to $\chi(0)$ gives
\begin{equation}
\label{chi0md6}
\chi(0)=\sqrt{\frac{140\alpha'm_\infty^2}{3\kappa}+\frac{49}{288}(6-d)^2}-\frac{7\sqrt{2}}{24}(6-d)\ .
\end{equation}
Note that this result was also derived in~\cite{Balthazar:2022hno}, which followed a slightly different method. However, the approach here should essentially be equivalent to the one used there.

For $d<6$, from (\ref{chi0md6}), we find the following approximations for $\chi(0)$:

1. When \(\alpha'm_\infty^2 \ll 6 - d \):
\begin{equation}
\label{dle61}
\chi(0)=\frac{40\sqrt{2}\alpha'm_\infty^2}{\kappa(6-d)}\ .
\end{equation}

2. When \( 6 - d \ll \alpha'm_\infty^2 \):
\begin{equation}
\label{dle62}
\chi(0)=\sqrt{\frac{140\alpha'}{3\kappa}}m_\infty\ .
\end{equation}
From (\ref{Mchi0}), along with (\ref{dle61}), (\ref{dle62}) and $\alpha' m_\infty^2\ll 1$, we see that mass lies in the range
\begin{equation}
\label{Mreg}
    M\gg M_c\ ,
\end{equation}
which is the same as the regime of validity for the Schwarzschild black hole.

On the other hand, for $d$ close to 6 from above, the expansion of (\ref{chi0md6}) around $m_\infty=0$ and $d=6$ yields
\begin{equation}
\label{dge61}
\chi(0)=\frac{7\sqrt{2}}{12}(d-6)+\frac{40\sqrt{2}\alpha'm_\infty^2}{\kappa(d-6)}\ ,\quad \alpha'm_\infty^2\ll d-6\ ,
\end{equation}
where we include the subleading term to show the dependence on $m_\infty$, and
\begin{equation}
\label{dge62}
\chi(0)=\sqrt{\frac{140\alpha'}{3\kappa}}m_\infty\ ,\quad d-6\ll \alpha'm_\infty^2
\end{equation}
which is the same as (\ref{dle62}). Substituting (\ref{dge61}) and (\ref{dge62}) into (\ref{Mchi0}), with $\alpha'm_\infty^2\ll 1$, leads to the mass range given by
\begin{equation}
\label{Mreg2}
    M_c\ll M\lesssim \frac{M_c}{d-6}\ .
\end{equation}
In comparison with (\ref{Mreg}), here is an upper bound of order $M_c/(d-6)$, which goes to infinity as $d\to 6^+$.

Also note that in the leading order, the free energy of the solution (\ref{d6HP}) is given by the first term in (\ref{Idmd}), namely
\begin{equation}
\label{Fd6}
    F=\frac{72\omega_5}{5\pi G_N^{(7)}}\frac{\alpha'^2}{\kappa^2}\ .
\end{equation}
It remains large as long as the string coupling is small, and thus quantum effects are suppressed.

We then proceed to calculate the entropy. In the leading order, we have
\begin{equation}
\label{SHP6}
\begin{split}
    S=&\beta_HM+(\beta-\beta_H)M-\beta_HF\\
    =&\beta_HM+\frac{2(\pi\alpha')^2}{\beta_H}m_\infty^2M-\beta_H\frac{72\omega_5}{5\pi G_N^{(7)}}\frac{\alpha'^2}{\kappa^2}\ .
\end{split}
\end{equation}

We now show that the second term in (\ref{SHP6}) is always subleading compared to the third term. For the case of $\alpha'm_\infty^2\ll 6-d$, we can express $m_\infty^2$ in terms of $M$ by using (\ref{dle61}) and (\ref{Mchi0}),
\begin{equation}
    m_\infty^2=\frac{36(6-d)\omega_5}{5\pi G_N^{(7)}M}\frac{\alpha'}{\kappa}\ .
\end{equation}
Plugging it in (\ref{SHP6}), one can easily see that the second term is of a higher order in $d-6$ than the third term. Therefore, this term can be neglected and the entropy simplifies to
\begin{equation}
\label{SHPl6}
    S=\beta_HM-\beta_H\frac{72\omega_5}{5\pi G_N^{(7)}}\frac{\alpha'^2}{\kappa^2}\ .
\end{equation}
For the case of $\alpha'm_\infty^2\ll d-6$, using (\ref{Mchi0}) and (\ref{dge61}), we find that the second term in (\ref{SHP6}) is of the order $m_\infty^2/(d-6)$, which is negligible compared to the third term. Thus, we can again express the entropy as in (\ref{SHPl6}).

For the case of $|d-6|\ll \alpha'm_\infty^2$, from (\ref{dle62}) and (\ref{Mchi0}), we find
\begin{equation}
\label{mGM}
    m_\infty^2=\frac{124416\omega_5^2}{35\pi^2\left(G_N^{(7)}M\right)^2}\frac{\alpha'^3}{\kappa^3}\ .
\end{equation}
Substituting it into (\ref{SHP6}) gives us
\begin{equation}
\label{SHPl61}
    S=\beta_HM+\frac{1}{\beta_H}\frac{248832\omega_5^2}{35\left(G_N^{(7)}\right)^2M}\frac{\alpha'^5}{\kappa^3}-\beta_H\frac{72\omega_5}{5\pi G_N^{(7)}}\frac{\alpha'^2}{\kappa^2}\ .
\end{equation}
However, as $M\gg M_c$, the second term is much smaller than the third term. Thus, the entropy formula again reduces to (\ref{SHPl6}).
\begin{figure}
	\centering
	\subfigure[]{
	\begin{minipage}[t]{0.45\linewidth}
	\centering
	\includegraphics[width=2.5in]{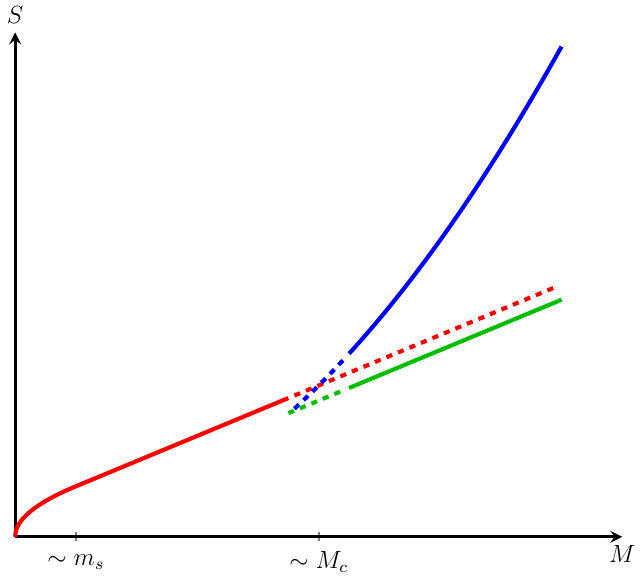}\label{entropiesdl6}
	\end{minipage}}
	\subfigure[]{
	\begin{minipage}[t]{0.45\linewidth}
	\centering
	\includegraphics[width=2.5in]{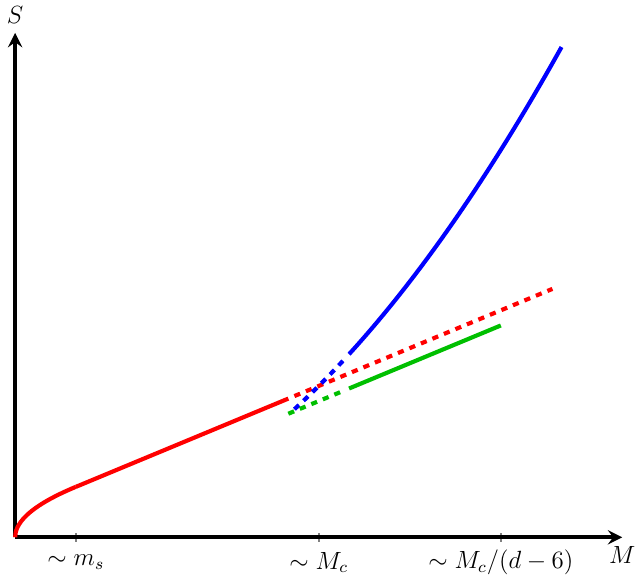}\label{entropiesdg6}
	\end{minipage}}
	\centering
\caption{\label{entropiesd6}Entropies of the Schwarzschild black hole (\ref{SBH}), the string star (\ref{SHPl6}) and the free string (\ref{Sfree}), denoted by blue, green, red lines respectively, for $d$ close to 6 (a) from below and (b) from above.}
\end{figure}

In summary, for $d$ close to 6, the string star has an entropy as in (\ref{SHPl6}). The regime of validity is given in (\ref{Mreg}) or (\ref{Mreg2}), depending whether $d$ is smaller or larger than 6. Note that the black hole solution also exists when $M\gg M_c$, and in this regime, the black hole entropy dominates over that of the string star. This indicates that the self-gravitation of the fundamental string is negligible during the transition from the black hole to the string state, as expected. The entropies of the three states are sketched in figure \ref{entropiesd6}. Note that in principle there is an additional scale $M_g$ in figure \ref{entropiesdl6}, similar to what is shown in figure \ref{entropiesdg4}. However, based on the definition in (\ref{Mg}), it is clear that $M_g$ is non-perturbative in $d-6$. And for $d$ close to 6 from below, $M_g$ is infinite at leading order, pushing that scale far to the right in figure \ref{entropiesdl6}. Furthermore, for $d$ close to 6 from above, the upper bound of $M$, which is in the order of $M_c/(d-6)$, corresponds to the solution at exactly $m_\infty=0$. Therefore, in figure \ref{entropiesdg6} we do not extrapolate the Horowitz-Polchinski solution beyond that point.
\section{Black string and Gregory-Laflamme instability}\label{secGL}
In the previous section, we have reviewed the transition between black holes and fundamental strings in $d+1$ noncompact dimensions. We now consider the $(d+2)$-dimensional spacetime asymptotic to $\mathbb{R}^{d,1}\times \mathbb{S}^1$, where the circumference of the spatial circle is denoted by $L$. When $L$ is sufficiently small, presumably the extra circle can be ignored, reducing the problem to the $(d+1)$-dimensional noncompact spacetime discussed in the previous section. As $L$ increases, however, a richer structure may emerge, including new solutions which are not uniform along the compact dimension. From now on, we will focus on the case where $L\gg l_s$, ensuring that the variation of non-uniform solutions in the spatial circle is small so that $\alpha'$ corrections are suppressed.

First, in the presence of an extra spatial circle, denoted by $z$, a trivial extension of the Schwarzschild metric (\ref{schw}) is
\begin{equation}
\label{bsmet}
    ds^2=-\left(1-\left(\frac{r_H}{r}\right)^{d-2}\right)dt^2+\frac{dr^2}{1-\left(\frac{r_H}{r}\right)^{d-2}}+r^2d\Omega_{d-1}^2+dz^2\ .
\end{equation}
Namely, the $z$ circle acts as a spectator. The horizon, which is at $r=r_H$, has a topology of $\mathbb{S}^{d-1}\times \mathbb{S}^1$, with the $\mathbb{S}^1$ factor indicating that the horizon wraps around the $z$ circle. Due to this additional factor, the solution (\ref{bsmet}) is called a black string. In contrast, we will preserve the term ``black hole'' for solutions with a spherical horizon topology $\mathbb{S}^d$. In other words, instead of wrapping around the $z$ circle, a black hole is localized in the $z$ dimension.

The mass and entropy of the black string (\ref{bsmet}) are still given by (\ref{M}) and (\ref{S}), respectively. And the entropy-mass relation follows (\ref{SBH}) with no change. Note that the $(d+1)$-dimensional Newton's constant is related to the $(d+2)$-dimensional one via
\begin{equation}
\label{GN}
    G_N^{(d+2)}=G_N^{(d+1)}L\ .
\end{equation}

The uniform black string (\ref{bsmet}) may encounter an instability towards non-uniformity along the $z$ direction, known as the Gregory-Laflamme instability~\cite{Gregory:1993vy,Gregory:1994bj}. In order to show this instability, one needs to find a perturbed solution around the background (\ref{bsmet}) with the lowest non-zero wavenumber in $z$, i.e. $k_z= 2\pi/L$, that grows in time as $e^{\Omega t}$, with some $\Omega>0$. Specifically, this can be done by perturbing the metric
\begin{equation}
    g_{\mu\nu}\to g_{\mu\nu}+\delta g_{\mu\nu}\ ,
\end{equation}
with $\delta g_{\mu\nu}\sim e^{\Omega t}\cos k_z z$, under which the Ricci tensor acquires a small change. At leading (linear) order,
\begin{equation}
    R_{\mu\nu}\to R_{\mu\nu}-\frac{1}{2}\Delta^{(d+2)}_L\delta g_{\mu\nu}\ .
\end{equation}
Here $\Delta^{(d+2)}_L$ is the $(d+2)$-dimensional Lichnerowicz operator. The specific expression of this operator is not important here. Since the Einstein's equations require $R_{\mu\nu}=0$, the perturbation $\delta g_{\mu\nu}$ to the metric must satisfy the following linear differential equations,
\begin{equation}
\label{DLg}
    \Delta^{(d+2)}_L\delta g_{\mu\nu}=0\ .
\end{equation}

It turns out that for (\ref{DLg}) to have solutions with real $\Omega$, $r_H$ has to be smaller than a critical value $r_{H,\text{GL}}$~\cite{Gregory:1993vy,Gregory:1994bj}. Therefore, when $r_H>r_{H,\text{GL}}$, the black string is stable against non-uniformity. The critical radius $r_{H,\text{GL}}$ is proportional to $L$, the only dimensionful parameter other than $r_H$ in Einstein's equations. Since we are focusing on the case where $L\gg l_s$, $r_{H,\text{GL}}$ is much larger than $l_s$. Thus, the instability occurs at a scale where $\alpha'$ corrections (outside the horizon) are still negligible. 

At the critical point $r_{H,\text{GL}}$, the black string undergoes a phase transition and becomes either a non-uniform black string (with horizon topology $\mathbb{S}^{d-1}\times \mathbb{S}^1$) or a black hole (with horizon topology $\mathbb{S}^d$), depending on the number of dimensions (for more details, see~\cite{Emparan:2018bmi}). Moreover, when this phase transition occurs, the entropy changes discontinuously for $d\lesssim 11.5$, but changes continuously for $d\gtrsim 11.5$~\cite{Sorkin:2004qq}. Thus, the phase transition is first-order for $d\lesssim 11.5$ and second-order for $d\gtrsim 11.5$.

In general, there are no analytic results for the non-uniform solution, and one has to rely on numerical methods. However, for sufficiently small mass, the solution can be well approximated by a higher-dimensional Schwarzschild black hole localized in the compact dimension. It has a horizon with the topology of a $d$-dimensional sphere and a radius of $r_H$. This approximation is valid when the size of the extra dimension is much larger than the size of the black hole, i.e. $L\gg r_H$. In this case, the $M$-$r_H$ relation follows the one in (\ref{M}), with $d$ replaced by $d+1$. More explicitly,
\begin{equation}
\label{Md1}
    M=\frac{d\ \omega_d}{16\pi G_N^{(d+1)}L}r_H^{d-1}\ ,
\end{equation}
where we have expressed $G_N^{(d+2)}$ in terms of $G_N^{(d+1)}$ using (\ref{GN}). Due to $L\gg r_H$, the mass (\ref{Md1}) is much smaller than the critical mass at $r_{H,\text{GL}}$, defined by
\begin{equation}
\label{MBSGL}
    M_\text{BS,GL}\equiv\frac{(d-1)\omega_{d-1}}{16\pi G_N^{(d+1)}}r_{H,\text{GL}}^{d-2}\sim \frac{L^{d-2}}{G_N^{(d+1)}}\ .
\end{equation}
Besides, the gravity result breaks down when the horizon radius is of order $l_s$, or
\begin{equation}
    M\sim \frac{l_s^{d-1}}{ G_N^{(d+2)}}=
    \frac{M_cl_s}{L}\ ,
\end{equation}
with $M_c$ given by (\ref{Mc}).
\begin{figure}
	\centering
	\subfigure[]{
	\begin{minipage}[t]{0.45\linewidth}
	\centering
	\includegraphics[width=2.5in]{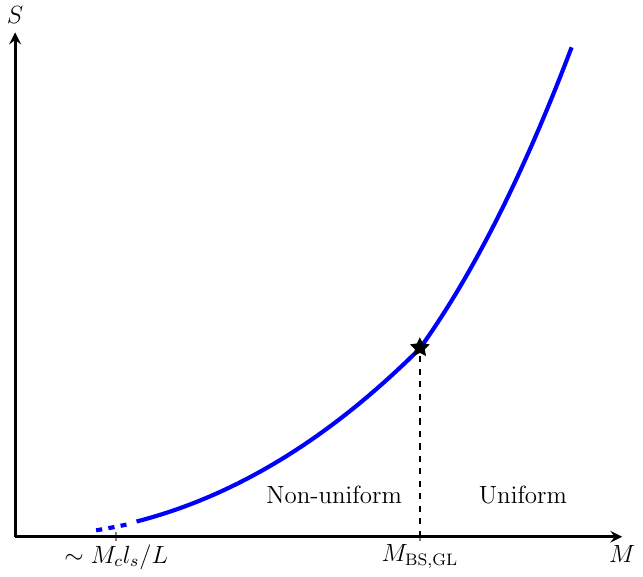}\label{entropiesBS}
	\end{minipage}}
	\subfigure[]{
	\begin{minipage}[t]{0.45\linewidth}
	\centering
	\includegraphics[width=2.5in]{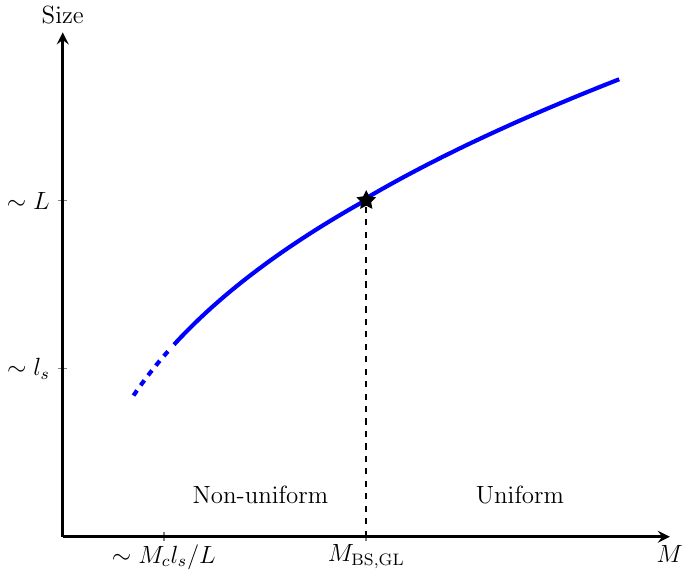}\label{sizeBS}
	\end{minipage}}
	\centering
\caption{\label{figureBS}Entropy and size of the black string (or black hole) in $\mathbb{R}^{d,1}\times \mathbb{S}^1$. The star marks the critical point where the Gregory-Laflamme instability occurs. The uniform black string corresponds to the portion of the solid blue at $M>M_\text{BS,GL}$, while the non-uniform solution (either a black string or a localized black hole) is found at $M<M_\text{BS,GL}$. The entropy is (dis)continuous at the critical point for $d\gtrsim 11.5$ ($d\lesssim 11.5$). For $M\ll M_\text{BS,GL}$, the non-uniform solution is well approximated by the $(d+2)$-dimensional Schwarzschild black hole.}
\end{figure}

In summary, for $L\gg l_s$, as $M$ decreases, the uniform black string solution transitions into a non-uniform solution at $M_\text{BS,GL}$, which gradually approaches a higher-dimensional Schwarzschild black hole as the mass decreases further, well below $M_\text{BS,GL}$. The entropy-mass relation is sketched in figure \ref{entropiesBS}. One can also see the change of the size with respect to the mass as in figure \ref{sizeBS}.

\section{Non-uniformity of the string star}\label{secnHP}
In this section, we study the string star solutions in spacetime with an extra spatial circle, $\mathbb{S}^1_z$. To this end, as in section \ref{sechp}, we first Wick rotate the time coordinate $\tau=it$. Then, reducing on the Euclidean time circle we have the $\mathbb{R}^d\times\mathbb{S}^1_z$ analog of the Horowitz-Polchinski effective action (\ref{Sphichid}), i.e.,
\begin{equation}
\label{Sphichiz}
	\begin{split}
 	I_{d+1}=\frac{\beta}{16\pi G_N^{(d+2)}}\int_0^Ldz\int d^dx&\bigg[(\partial_z\varphi)^2+(\nabla_x\varphi)^2+|\partial_z \chi|^2+|\nabla_x \chi|^2\\
  &+\left(m_\infty^2+\frac{\kappa}{\alpha'}\varphi\right)|\chi|^2\bigg]\ .
	\end{split}
	\end{equation}
 
If the fields $\varphi$ and $\chi$ are independent of $z$, we can further integrate over $z$ in (\ref{Sphichiz}) and recover the original Horowitz-Polchinski effective action on $\mathbb{R}^d$, (\ref{Sphichid}). This leads to the uniform Horowitz-Polchinski solution where $z$ acts as a spectator. This solution describes a uniform string star in Lorentzian signature.\footnote{Note that the Horowitz-Polchinski solution does not directly give rise to any physical state when we Wick rotate back to the Lorentzian signature; after all, the Lorentzian time is not periodic, making it nonsensical to discuss the winding tachyon as in the Euclidean signature. We stress that it describes a string star in the sense that it yields the correct thermodynamic quantities, such as mass and entropy.} In contrast, a non-uniform solution of (\ref{Sphichiz}), which depends on the $z$ coordinate, describes a string star that is non-uniform along the $z$ circle.

Drawing from the black string case, we learn that a uniform solution becomes unstable towards non-uniformity when its size is small compared to the length of the spatial circle. Therefore, for the uniform string star, we anticipate a Gregory-Laflamme instability when its characteristic length—proportional to $m_\infty^{-1}$ according to the rescaling property in (\ref{resHP})—becomes smaller than $L$ up to some factor.

Note that discussing the Gregory-Laflamme instability of the string star using the Horowitz-Polchinski effective field theory involves a subtlety. As in the black string case, genuinely demonstrating the Gregory-Laflamme instability of a uniform solution requires identifying a time-dependent, growing non-uniform perturbation. However, in the effective field theory (\ref{Sphichiz}), we are working in Euclidean signature, which precludes any discussion of time-dependent solutions. Therefore, instead of addressing the dynamic instability, we will focus on the thermodynamic instability—an instability signaled by a decrease in free energy under a perturbation. Specifically, as we will see in the following subsection, when $m_\infty$ exceeds a critical value proportional to $1/L$, the free energy decreases under a non-uniform perturbation, signaling an instability towards non-uniformity. We acknowledge that we are assuming the equivalence of dynamic and thermodynamic instabilities, known as the Gubser-Mitra conjecture~\cite{Gubser:2000ec,Gubser:2000mm}, which may not be true in general (see, e.g.,~\cite{Bostock:2004mg,Marolf:2004fya,Friess:2005zp}). We do not attempt to address this issue in this paper. And with some abuse of terminology, we refer to this thermodynamic instability as the Gregory-Laflamme instability of the string star.

\subsection{Critical point: leading-order perturbations $\delta\chi^{(1)}$ and $\delta\varphi^{(1)}$}\label{secHPLO}
We notice in the previous section that at the critical horizon radius $r_{H,\text{GL}}$ of the uniform black string, the leading-order non-uniform perturbation $\delta g_{\mu\nu}$ exhibits $\Omega=0$, indicating that the solution is time-independent. This implies that by solving the e.o.m. in Euclidean signature for a $\tau$-independent perturbation with $k_z=2\pi/L$, we can identify the critical point, which marks the onset of the Gregory-Laflamme instability. In this subsection, we employ this strategy to determine the critical point for the Gregory-Laflamme instability of the uniform string star.

Let us denote the uniform Horowitz-Polchinski solutions, namely the solutions of (\ref{eomHP}), as $\chi_*(r)$ and $\varphi_*(r)$. We treat them as background solutions and introduce non-uniform perturbations $\delta\chi$ and $\delta\varphi$. To leading order, the perturbations take the form 
\begin{equation}
\label{chiphi1}
    \delta\chi^{(1)}(r,z)=\lambda\chi_1(r)\cos k_z z\ ,\quad \delta\varphi^{(1)}(r,z)=\lambda\varphi_1(r)\cos k_z z\ ,
\end{equation}
with $k_z=2\pi/L$ and $\lambda$ is introduced as a perturbative parameter. 

To find the leading-order perturbations (\ref{chiphi1}), we first substitute the perturbed fields
\begin{equation}
\label{chiphiper}
    \chi(r,z)=\chi_*(r)+ \delta\chi^{(1)}(r,z)\ ,\ \varphi(r,z)=\varphi_*(r)+\delta\varphi^{(1)}(r,z)
\end{equation}
into the action (\ref{Sphichiz}) and integrate over $z$. This procedure yields a $d$-dimensional effective action $\chi_1$ and $ \varphi_1$, which reads
\begin{equation}
	\label{Sphichizdel}
	\begin{split}
 	\delta I_{d}=\frac{\lambda^2\beta}{32\pi G_N^{(d+1)}}\int d^dx\left[(\nabla\varphi_1)^2+(\nabla \chi_1)^2+\left(m_{\infty}^2+k_z^2+\frac{\kappa}{\alpha'}\varphi_*\right)\chi^2_1+k_z^2\varphi^2_1+\frac{2\kappa}{\alpha'}\chi_*\varphi_1\chi_1\right]\ .
	\end{split}
	\end{equation}
 
The action (\ref{Sphichizdel}) leads to the following equations of motion for spherically symmetric solutions $\chi_1(r)$ and $\varphi_1(r)$,
        \begin{equation}
        \label{eomper}
	\begin{split}
 	\chi_1''(r)+\frac{d-1}{r}\chi_1'(r) &=\left(m_{\infty}^2+k_z^2+\frac{\kappa}{\alpha'}\varphi_*(r)\right)\chi_1(r)+\frac{\kappa}{\alpha'}\chi_*(r)\varphi_1(r) \ ,\\
 	\varphi_1''(r)+\frac{d-1}{r}\varphi_1'(r) &=k_z^2\varphi_1(r)+\frac{\kappa}{\alpha'}\chi_*(r)\chi_1(r)\ .
	\end{split}
	\end{equation}
The normalizable solutions of these equations satisfy the boundary conditions that $\chi_1$ and $\varphi_1$ remain finite at $r=0$ and approach zero as $r\to \infty$. These conditions determine the critical value $m_{\infty,\text{GL}}$ for us, analogous to how the normalization condition determines the eigenvalues of the Schrodinger equation in quantum mechanics.
\begin{figure}
	\centering
	\subfigure[]{
	\begin{minipage}[t]{0.45\linewidth}
	\centering
	\includegraphics[width=2.5in]{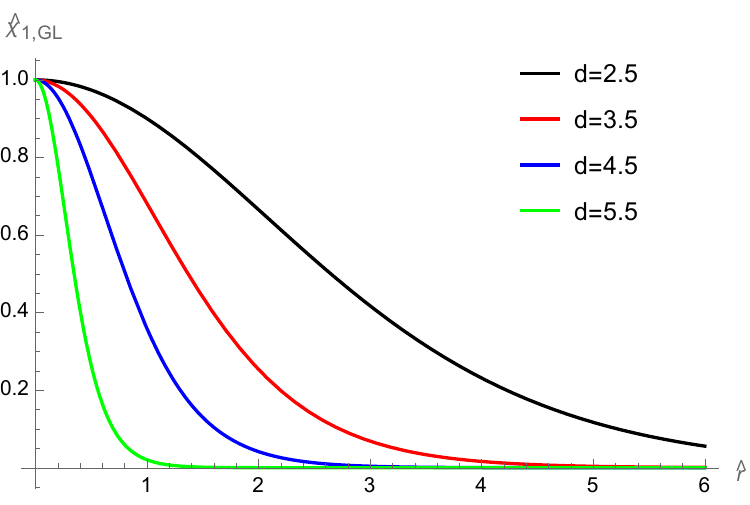}\label{delchi1}
	\end{minipage}}
	\subfigure[]{
	\begin{minipage}[t]{0.45\linewidth}
	\centering
	\includegraphics[width=2.5in]{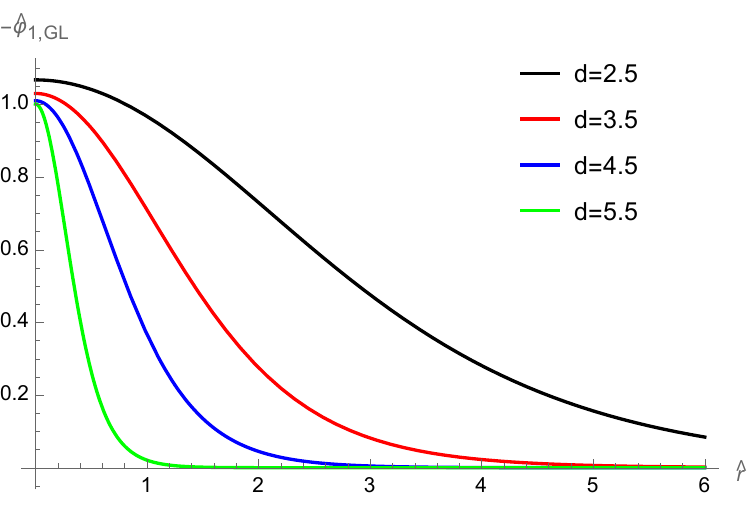}\label{delphi1}
	\end{minipage}}
	\centering
\caption{\label{del1}Profiles of the leading-order perturbations $\hat\chi_{1,\text{GL}}$ and $-\hat\varphi_{1,\text{GL}}$ for $d=2.5$, 3.5, 4.5 and 5.5.}
\end{figure}

To solve the e.o.m., it is again convenient to perform the rescaling (\ref{resHP}), which transforms (\ref{eomper}) into
\begin{equation}
\label{eomperres}
	\begin{split}
 	\hat\chi_1''(\hat{r})+\frac{d-1}{\hat{r}}\hat\chi_1'(\hat{r}) &=\left(1+\hat{k}_z^2+\hat{\varphi}_*(\hat r)\right)\hat\chi_1(\hat r)+\hat\chi_*(\hat r)\hat\varphi_1(\hat r) \ ,\\
 	\hat\varphi_1''(\hat r)+\frac{d-1}{\hat r}\hat\varphi_1'(\hat r) &=\hat k_z^2\hat\varphi_1(\hat r)+ 2\hat\chi_*(\hat r)\hat\chi_1(\hat r)\ .
	\end{split}
	\end{equation}
Here, 
\begin{equation}
\label{hatkz}
    \hat k_z\equiv \frac{k_z}{m_\infty}\ .
\end{equation}

We can solve (\ref{eomperres}) numerically. As they are linear, the amplitude of the solutions is arbitrary, and we use this freedom to set $\hat\chi_1(0)=1$. We then apply the shooting method to numerically solve the differential equations, adjusting the values of $\hat k_z$ and $\hat\varphi_1(0)$ until we obtain a solution where both $\hat\chi_1$ and $\hat\varphi_1$ vanish asymptotically. The normalizable solutions, found through this process will be denoted by $\hat\chi_{1,\text{GL}}$ and $\hat\varphi_{1,\text{GL}}$, while the corresponding value of $\hat k_z$ is labeled $\hat k_\text{GL}$. 

In figure \ref{del1}, we plot the numerical solutions $\hat\chi_{1,\text{GL}}$ and $\hat\varphi_{1,\text{GL}}$ that we find for $d=2.5$, 3.5, 4.5 and 5.5. The values of $\hat k_\text{GL}$ are determined to be 0.5415, 1.025, 1.667 and 3.654, respectively. We also find that as $d$ approaches 6, $\hat k_\text{GL}$ tends to diverge to infinity.

Now, from the definition of $\hat k_z$, (\ref{hatkz}), the critical value of $m_\infty$ can be determined as 
\begin{equation}
\label{mGL}
    m_{\infty,\text{GL}}=\frac{k_z}{\hat k_\text{GL}}=\frac{2\pi}{\hat k_\text{GL}L}\ .
\end{equation}
Since $\hat k_\text{GL}$ is an order 1 number for $d$ strictly smaller than 6 and we are focusing on the regime where $L\gg l_s$, we always have $\alpha'm_{\infty,\text{GL}}^2\ll 1$. Therefore, the Horowitz-Polchinski effective action (\ref{Sphichiz}) remains valid at the critical point. Besides, we want to assume that quantum effects can be neglected at the critical point. This leads to the condition $I_d(m_{\infty,\text{GL}})\gg 1$. From (\ref{MhatC}) and (\ref{FId}), we find that $I_d(m_{\infty,\text{GL}})\sim m_{\infty,\text{GL}}^{6-d}l_s^5/G_N^{(d+1)}$. Thus, this condition imposes the constraint
\begin{equation}
\label{GLcon}
    L\ll \left(\frac{l_s^5}{G_N^{(d+1)}}\right)^{\frac{1}{6-d}}\ .
\end{equation}

Finally, we can use the solutions $\chi_{1,\text{GL}}(r)$ and $\varphi_{1,\text{GL}}(r)$ to examine the thermodynamic instability towards non-uniformity for $m_\infty>m_{\infty,\text{GL}}$. In particular, at a given $m_\infty$, substituting these solutions into the action (\ref{Sphichizdel}) yields the following change in free energy,
\begin{equation}
\label{freeGL}
	\begin{split}
 	\delta F=&\frac{\lambda^2\omega_{d-1}}{32\pi G_N^{(d+1)}}\int dr\ r^{d-1}\bigg[-\varphi_{1,\text{GL}}\nabla^2\varphi_{1,\text{GL}}-\chi_{1,\text{GL}}\nabla^2\chi_{1,\text{GL}}\\
  &+\left(m_{\infty}^2+k_z^2+\frac{\kappa}{\alpha'}\varphi_*|_{m_\infty}\right)\chi^2_{1,\text{GL}}+k_z^2\varphi^2_{1,\text{GL}}+\frac{2\kappa}{\alpha'}\chi_*|_{m_\infty} \varphi_{1,\text{GL}}\chi_{1,\text{GL}}\bigg]\ ,
	\end{split}
	\end{equation}
where we have integrated (\ref{Sphichizdel}) by parts and divided the result by $\beta$. Since $\chi_{1,\text{GL}}$ and $\varphi_{1,\text{GL}}$ satisfy the equations (\ref{eomper}) \textit{at} $m_{\infty,\text{GL}}$, this simplifies to
\begin{equation}
\label{freeGLsim}
	\begin{split}
 	\delta F=\frac{\lambda^2\omega_{d-1}}{32 \pi G_N^{(d+1)}}\int dr\ r^{d-1}\bigg[&(m_\infty^2-m_{\infty,\text{GL}}^2+\frac{\kappa}{\alpha'}\varphi_*|_{m_\infty}-\frac{\kappa}{\alpha'}\varphi_*|_{m_{\infty,\text{GL}}} )\chi^2_{1,\text{GL}}\\
  &+(\frac{2\kappa}{\alpha'}\chi_*|_{m_\infty}-\frac{2\kappa}{\alpha'}\chi_*|_{m_{\infty,\text{GL}}})\varphi_{1,\text{GL}}\chi_{1,\text{GL}}\bigg]\ .
\end{split}
\end{equation}
By performing the rescaling (\ref{resHP}) and defining
\begin{equation}
\label{f1}
\begin{split}
    f_1\equiv\int d\hat r\ \hat r^{d-1}\bigg[(1+\hat\varphi_*) \hat\chi^2_{1,\text{GL}}+2\hat\chi_*\hat\varphi_{1,\text{GL}}\hat\chi_{1,\text{GL}}\bigg]\ ,
\end{split}
\end{equation}
we can rewrite (\ref{freeGLsim}) as
\begin{equation}
\label{delF}
\delta F =\lambda^2\frac{(m_\infty^2-m_{\infty,\text{GL}}^2)m_\infty^{4-d}\omega_{d-1}}{16 \pi G_N^{(d+1)}}\left(\frac{\alpha'}{\kappa}\right)^2 f_1\ .
	\end{equation}

The value of $f_1$ is calculated numerically, as shown in table \ref{coes} for various values of $d$. We see that it is always negative. Thus, when $m>m_{\infty,\text{GL}}$, it follows from (\ref{delF}) that $\delta F<0$. This indicates an instability towards non-uniformity, or a Gregory-Laflamme instability. 

\subsection{Order of phase transition: second-order perturbations $\delta\chi^{(2)}$ and $\delta\varphi^{(2)}$}\label{secHPNLO}
\begin{figure}
	\centering
	\subfigure[]{
	\begin{minipage}[t]{0.45\linewidth}
	\centering
	\includegraphics[width=2.5in]{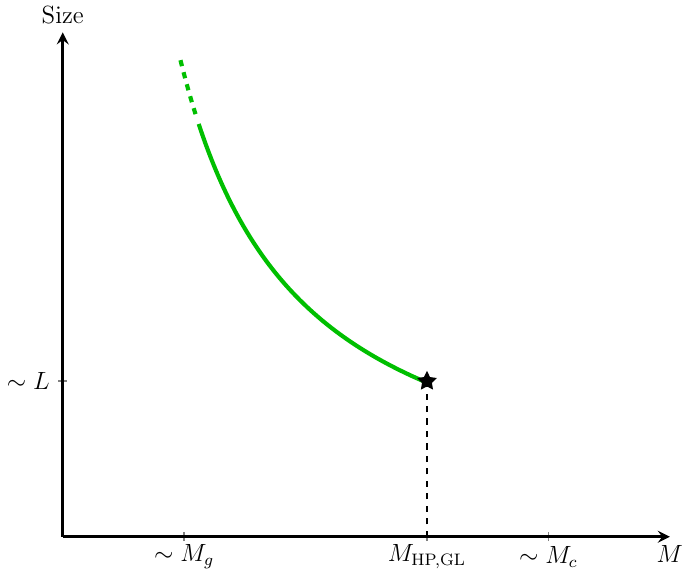}\label{sizeHPdl4}
	\end{minipage}}
	\subfigure[]{
	\begin{minipage}[t]{0.45\linewidth}
	\centering
	\includegraphics[width=2.5in]{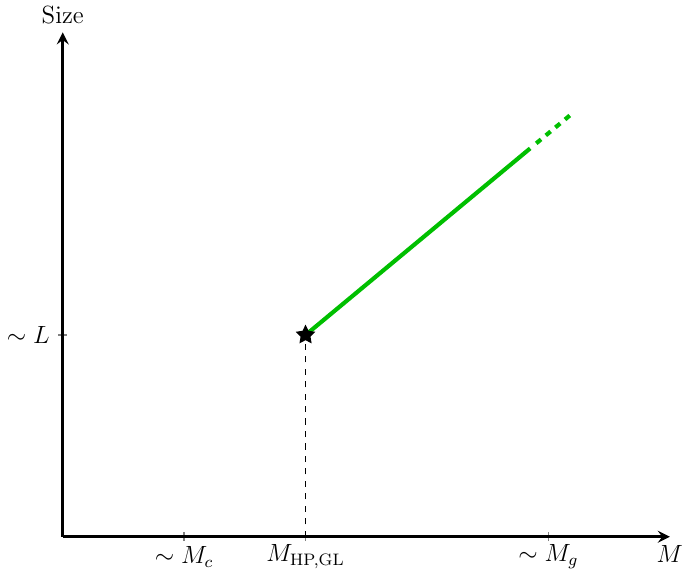}\label{sizeHPdg4}
	\end{minipage}}
	\centering
\caption{\label{sizeHP}Size of the string star in the uniform region (\ref{uniform}) for (a) $d<4$ and (b) $4<d<6$}
\end{figure}
In this subsection, we further investigate the order of the phase transition for the microcanonical ensemble at the critical point of the Gregory-Laflamme instability. First, in the preceding analysis, we established that when $m_\infty<m_{\infty,\text{GL}}$, the uniform string star is stable against non-uniformity. Using the relation between $m_\infty$ and the mass $M$ as given in (\ref{MhatC}), this condition defines a uniform region for the microcanonical ensemble, i.e.
\begin{equation} 
\label{uniform} 
M < M_{\text{HP,GL}} \text{ for } d < 4, \quad M > M_{\text{HP,GL}} \text{ for } 4 < d < 6\ . 
\end{equation}
Here, $M_\text{HP,GL}$ represents the mass at $m_{\infty,\text{GL}}$. From the expression of $m_{\infty,\text{GL}}$, (\ref{mGL}), we find
\begin{equation}
\label{MHPGL}
    M_\text{HP,GL}\sim \frac{\alpha'L^{d-4}}{G_N^{(d+1)}}\ .
\end{equation}
In figure \ref{sizeHP}, we sketch the size of the uniform solution with respect to the mass in the region (\ref{uniform}). It can be seen that as $M$ approaches $M_\text{HP,GL}$, the size becomes comparable to $L$.

The order of the phase transition for the microcanonical ensemble depends on whether the change in entropy is continuous or discontinuous at the critical point. To assess this, we evaluate the mass of the non-uniform solution perturbatively around the critical point. In particular, if the perturbative non-uniform solution has a mass within the uniform region defined in (\ref{uniform}), this suggests a first-order phase transition from the uniform solution to a non-perturbative non-uniform solution, accompanied by a discontinuous change of entropy. Conversely, if the perturbative non-uniform solution lies outside the uniform region, the entropy changes continuously during the phase transition, classifying it as second-order.

It is convenient to derive the mass using the canonical ensemble. First, when $m_\infty$ slightly deviates from the critical value $m_{\infty,\text{GL}}$, the non-uniform perturbation is dominated by $\chi_{1,\text{GL}}(r)$ and $\varphi_{1,\text{GL}}(r)$. Thus, provided that $|m_\infty-m_{\infty,\text{GL}}|/m_{\infty,\text{GL}}$ is sufficiently small, we can express the free energy as a combination of the unperturbed one $F_0(m_\infty^2)$ and the change $\delta F$ given by (\ref{delF}). This leads to
\begin{equation}
\label{Flambda}
    F(m_\infty^2,\lambda)=F_0(m_\infty^2)+\frac{m_\infty^2-m_{\infty,\text{GL}}^2}{m_{\infty,\text{GL}}^2}F_1\lambda^2+O\left(\lambda^4\right)\ ,\quad \left|\frac{m_\infty^2-m_{\infty,\text{GL}}^2}{m_{\infty,\text{GL}}^2}\right|\ll \lambda\ ,
\end{equation}
where 
\begin{equation}
\label{F1}
\begin{split}
    F_1=&\frac{m_{\infty,\text{GL}}^{6-d}\omega_{d-1}}{16 \pi G_N^{(d+1)}}\left(\frac{\alpha'}{\kappa}\right)^2f_1\ .    
\end{split}
\end{equation}
Note that the free energy is an even function of $\lambda$ due to the symmetry under $\lambda\leftrightarrow -\lambda$.\footnote{This can be seen by shifting $z\to z+L/2$, under which $\delta\chi^{(1)},\delta\varphi^{(1)}\sim \lambda \cos 2\pi z/L\to -\lambda \cos 2\pi z/L$.} This is why the remaining terms in (\ref{Flambda}) are of order $\lambda^4$, instead of $\lambda^3$.

The free energy (\ref{Flambda}) needs to be extremized by differentiating it with respect to $\lambda$ and equating the derivative to zero. Denoting the extremum as $\lambda_*$, we find that $\lambda_*=0$ when considering only the $\lambda^2$ term. To obtain a nonzero value for $\lambda_*$, it is necessary to also account for the $\lambda^4$ term in the free energy. Indeed, as we will see below, the nonzero value of $\lambda_*^2$ is of the same order as $(m_\infty^2-m_{\infty,\text{GL}}^2)/m_{\infty,\text{GL}}^2$, which means that the $\lambda_*^4$ term is in the same order as the $\lambda_*^2$ term, explaining why the $\lambda^4$ term cannot be neglected in the extremization process.

In appendix \ref{secla4}, we calculate the coefficient of the $\lambda^4$ term in the free energy at $m_{\infty,\text{GL}}$. This is achieved by first finding the next-to-leading-order perturbations, $\delta^{(2)}\chi$ and $\delta^{(2)}\varphi$. They are sourced by quadratic terms of the leading-order perturbations, such as $(\delta^{(1)}\chi)^2$. These quadratic terms are proportional to $\cos^2 k_zz$. By decomposing $\cos^2 k_zz$ as $1+\cos 2k_zz$, we see that the source terms couple to fields with wavenumbers 0 and $2k_z$. Thus, we can express the second order perturbations as
\begin{equation}
\label{del2}
\begin{split}
    \delta\chi^{(2)}(r,z)&=\lambda^2\left(\chi_0(r)+\chi_2(r)\cos 2k_zz\right)\ ,\\ \delta\varphi^{(2)}(r,z)&=\lambda^2\left(\varphi_0(r)+\varphi_2(r)\cos 2k_zz\right)\ .
\end{split}
\end{equation}

With the result derived in appendix \ref{secla4}, we can expand the free energy to include the $\lambda^4$ term (given by (\ref{delI})),
\begin{equation}
\label{Flambda4}
F(m_\infty^2,\lambda)=F_0(m_\infty^2)+\frac{m_\infty^2-m_{\infty,\text{GL}}^2}{m_{\infty,\text{GL}}^2}F_1\lambda^2+F_2\lambda^4+O\left(\lambda^6\right)\ ,\quad \left|\frac{m_\infty^2-m_{\infty,\text{GL}}^2}{m_{\infty,\text{GL}}^2}\right|\ll \lambda\ ,
\end{equation}
where
\begin{equation}
\label{F2}
\begin{split}
    F_2=&\frac{m_{\infty,\text{GL}}^{6-d}\omega_{d-1}}{16\pi G_N^{(d+1)}}\left(\frac{\alpha'}{\kappa}\right)^2f_2\ ,\\
    f_2=&\int d\hat r\ \hat r^{d-1}\bigg(\hat\chi_{1,\text{GL}}\hat\varphi_{1,\text{GL}}(\hat\chi_0+\frac{1}{2}\hat\chi_2)+\frac{1}{2}\hat\chi_{1,\text{GL}}^2(\hat\varphi_0+\frac{1}{2}\hat\varphi_2)\bigg)\ .
\end{split}
\end{equation}

The extremization of the free energy (\ref{Flambda4}) gives the (nontrivial) saddle points 
\begin{equation}
\label{lastar}
   \lambda_*^2=-\frac{m_\infty^2-m_{\infty,\text{GL}}^2}{2m_{\infty,\text{GL}}^2}\frac{F_1}{F_2}=-\frac{m_\infty^2-m_{\infty,\text{GL}}^2}{2m_{\infty,\text{GL}}^2}\frac{f_1}{f_2}\ .
\end{equation}
Note that from (\ref{F1}) and (\ref{F2}), the values of $f_1$ and $f_2$ are order 1, which can be determined numerically. This implies that $ (m_\infty^2-m_{\infty,\text{GL}}^2)/m_{\infty,\text{GL}}^2$ is in the same order as $\lambda_*^2$. So, for a small perturbative parameter, i.e. $\lambda_*\ll 1$, we have $ (m_\infty^2-m_{\infty,\text{GL}}^2)/m_{\infty,\text{GL}}^2\sim \lambda_*^2\ll \lambda_*$. This is consistent with the condition imposed in (\ref{Flambda4}), ensuring that the expansion remains valid for small perturbations. Also, from (\ref{lastar}), it is clear that whether the value of $m_\infty$ for the non-uniform perturbation is larger or smaller than $m_{\infty,\text{GL}}$ depends on the sign of $f_1/f_2$. If $f_1/f_2>0$, the non-uniform perturbations will result in a smaller $m_\infty$ compared to $m_{\infty,\text{GL}}$. Conversely, if $f_1/f_2<0$, the non-uniform perturbations will yield a larger $m_\infty$ than $m_{\infty,\text{GL}}$.

Substituting (\ref{lastar}) back into (\ref{Flambda4}), we can express the free energy in terms of $m_\infty$,
\begin{equation}
F(m_\infty^2)=F_0(m_\infty^2)-\frac{1}{4}\left(\frac{m_\infty^2-m_{\infty,\text{GL}}^2}{m_{\infty,\text{GL}}^2}\right)^2\frac{F_1^2}{F_2}+O\left(\left(\frac{m_\infty^2-m_{\infty,\text{GL}}^2}{m_{\infty,\text{GL}}^2}\right)^4\right)\ .
\end{equation}
And with this expression, we can now calculate the mass of the non-uniform perturbed solution.
\begin{equation}
\label{Mm1}
\begin{split}
M(m_\infty^2)=&\frac{d(\beta F)}{d\beta}\\
=&M_0(m_\infty^2) -\frac{\kappa}{2\alpha'}\frac{m_\infty^2-m_{\infty,\text{GL}}^2}{m_{\infty,\text{GL}}^4}\frac{F_1^2}{F_2}+O\left(\left(\frac{m_\infty^2-m_{\infty,\text{GL}}^2}{m_{\infty,\text{GL}}^2}\right)^2\right)\ ,    
\end{split}
\end{equation}
where $M_0(m_\infty^2)$ is the original mass before perturbations, i.e. (\ref{MhatC}). 

To determine whether (\ref{Mm1}) is larger or smaller than $M_\text{HP,GL}$, we expand $M_0$ around $m_{\infty,\text{GL}}$. This leads to the following form for the total mass.
\begin{equation}
\label{M0M1}
\begin{split}
M(m_\infty^2)=M_\text{HP,GL}+M_1(m_{\infty,\text{GL}}^2)\left(m_\infty^2-m_{\infty,\text{GL}}^2\right)+O\left(\left(\frac{m_\infty^2-m_{\infty,\text{GL}}^2}{m_{\infty,\text{GL}}^2}\right)^2\right)\ .
\end{split}
\end{equation}
The coefficient in the second term is given by
\begin{equation}
\begin{split}
M_1(m_{\infty,\text{GL}}^2)=&M_0'(m_{\infty,\text{GL}}^2) -\frac{\kappa}{2\alpha'}\frac{1}{m_{\infty,\text{GL}}^4}\frac{F_1^2}{F_2}\\ 
=&\frac{(d-2)(4-d)\omega_{d-1}}{16\pi G_N^{(d+1)}}\frac{\alpha'}{\kappa} m_{\infty,\text{GL}}^{2-d}\hat C_\varphi-\frac{\kappa}{2\alpha'}\frac{1}{m_{\infty,\text{GL}}^4}\frac{F_1^2}{F_2}\ ,
\end{split}
\end{equation}
where in the last step we have used (\ref{MhatC}). Using (\ref{F1}) and (\ref{F2}), we can further simplify the coefficient $M_1$ to
\begin{equation}
\label{M1m1}
    M_1=\frac{\alpha'}{\kappa}\frac{\omega_{d-1}m_{\infty,\text{GL}}^{2-d}}{32\pi G_N^{(d+1)}}m_1\ ,
\end{equation}
where
\begin{equation}
    m_1\equiv 2(d-2)(4-d)\hat C_\varphi-\frac{f_1^2}{f_2}\ .
\end{equation}
\begin{table}[t!]
\centering
\noindent
\hspace{-1.0cm}
\begin{tabular}{|c|c|c|c|c|c|c|}
 \hline\hline
  $d$ &2.5&3&3.5&4&4.5&5 \\  \hline
$\hat C_\varphi$   &2.110&3.505&5.397& 7.695& 9.827&10.40\\
$f_1  $ &-2.971& -3.196& -2.923& -2.225& -1.309&-0.5050\\
$f_2$ &0.9089&0.1388& 0.02201& 0.001066& -0.0007310&-0.0001817\\
$m_1$ &-6.549 & -66.58& -380.1& -4643& 2319&1341\\
\hline\hline
\end{tabular}
\caption[]{Numerical values of $\hat C_\varphi$, $f_1$, $f_2$ and $m_1$.}
\label{coes}
\end{table}

In table \ref{coes} we list the numerical values of $\hat C_\varphi$, $f_1$, $f_2$ and $m_1$ for several values of $d$. We notice that $f_2$ vanishes at a critical dimension $d_c$ slightly larger than 4; it is positive for $d<d_c$ and negative for $d>d_c$. This also explains why $m_1$ becomes large and negative as $d\to d_c^-$, and becomes large and positive as $d\to d_c^+$. For $d<d_c$, due to the positive $f_2$ and the negative $f_1$, we can see from (\ref{lastar}) that the perturbed non-uniform solutions are at $m_\infty>m_{\infty,\text{GL}}$. Furthermore, according to the numerical values in table \ref{coes}, $m_1$ is negative for $d<d_c$. Thus, in this case, using (\ref{M0M1}) and (\ref{M1m1}), we find that the mass decreases from $M_\text{HP,GL}$.

From the above analysis, we deduce that for $d<4$, the mass of the perturbed non-uniform solutions lies within the uniform region defined in (\ref{uniform}). As mentioned above, this indicates that the phase transition is first-order. In contrast, for $4<d<d_c$, the non-uniform perturbed solutions fall outside the uniform region, and thus the phase transition is second-order.

For the case of $d_c<d(<6)$, both $f_1$ and $f_2$ are negative, so the perturbed non-uniform solutions are at $m_\infty<m_{\infty,\text{GL}}$. Besides, $m_1>0$, which combining with (\ref{M0M1}) means that the mass decreases from $M_\text{HP,GL}$. In other words, the mass of the perturbed non-uniform solutions is out of the uniform region (\ref{uniform}). Therefore, the phase transition is second-order.

To summarize, in the microcanonical ensemble, the phase transition from the uniform string star to a non-uniform solution is first-order at $d<4$, while it is second-order at $4<d(<6)$.
\subsection{On non-uniform solutions}\label{secons}
Outside the region (\ref{uniform}), the solution becomes non-uniform along the $z$ circle. A key question now arises: how does the non-uniform solution behave, especially when the mass significantly deviates from the critical value $M_\text{HP,GL}$? In the previous section, we learned that for the black string case, when $M\ll M_\text{BS,GL}$, the non-uniform solution can be well approximated by the higher-dimensional Schwarzschild black hole localized in the compact dimension. This raises the question of whether a similar approximation holds for the string star. Specifically, as $M$ deviates from the critical value $M_\text{HP,GL}$ into the non-uniform region, does the solution approach the higher-dimensional string star localized in $\mathbb{S}_z^1$? In the following discussion, we will explore this possibility. For convenience, we denote the Horowitz-Polchinski solution derived from the noncompact $d$-dimensional effective action (\ref{Sphichid}) (or the uniform solution from (\ref{Sphichiz})) as HP$_d$. Thus, HP$_{d+1}$ can describe a higher-dimensional string star localized in the compact dimension. Besides, we maintain the notation for the $(d+1)$-dimensional string coupling, $g_s$, which remains related to $G_N^{(d+1)}$ as in (\ref{GN}).
\subsubsection{$2<d<3$}
We begin by discussing the case of $d<3$. Recall that the regime of validity for HP$_d$ is $M_g\ll M\ll M_c$, where $M_c$ and $M_g$ are defined by (\ref{Mc}) and (\ref{Mg}), respectively. Similarly, the regime of validity for HP$_{d+1}$ is $M_g^{(d+1)}\ll M\ll M_c^{(d+1)}$, with $M_c^{(d+1)}$ and $M_g^{(d+1)}$ defined analogously to (\ref{Mc}) and (\ref{Mg}), but with $d$ replaced by $d+1$, and $g_s^2$ replaced by $g_s^2L/l_s$. For clarity, we write these explicitly as follows:
\begin{equation}
\label{Mcgd1}
    M_c^{(d+1)}\equiv \frac{1}{g_s^2L}\ ,\quad M_g^{(d+1)}\equiv\left(\frac{g_s^4L^2}{l_s^{d-3}}\right)^{\frac{1}{d-5}}\ .
\end{equation}
Notably, using the condition (\ref{GLcon}), we find that 
\begin{equation}
\label{gsL}
    g_s^2L/l_s\ll g_s^{2(5-d)/(6-d)}\ll 1
\end{equation}
which leads to $M_g^{(d+1)}\ll  M_c^{(d+1)}$. Moreover, we have the following hierarchy of scales
\begin{equation}
\label{hi2d3}
    M_g\ll M_g^{(d+1)}\ll M_\text{HP,GL}\ll M_c^{(d+1)}\ll M_c\ ,\quad d<3\ ,
\end{equation}
where the first and second inequalities follow from (\ref{GLcon}), while the third and fourth inequalities follow from the condition $L\ll l_s$.

Outside the uniform region (\ref{uniform}), when the mass is much larger than $M_\text{HP,GL}$, it follows from (\ref{MhatC}) and (\ref{MHPGL}) that for HP$_{d+1}$, $m_\infty\gg 1/L$. This indicates that the length scale of HP$_{d+1}$ is much smaller than $L$, allowing it to fit well in the extra compact dimension.
\begin{figure}
	\centering
	\subfigure[]{
	\begin{minipage}[t]{0.45\linewidth}
	\centering
	\includegraphics[width=2.5in]{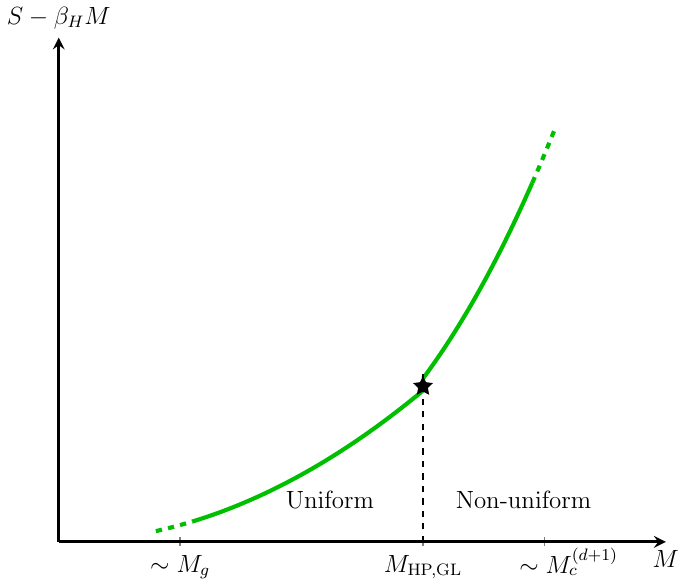}\label{entropiesHPd3}
	\end{minipage}}
	\subfigure[]{
	\begin{minipage}[t]{0.45\linewidth}
	\centering
	\includegraphics[width=2.5in]{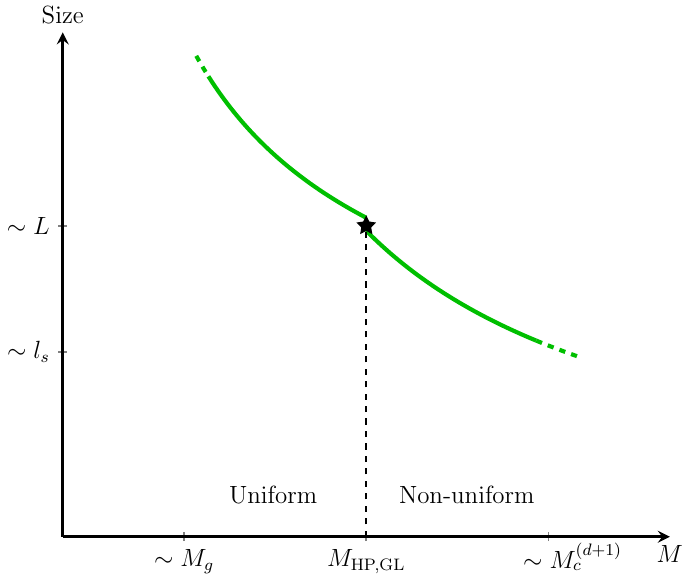}\label{sizeHPd3}
	\end{minipage}}
	\centering
\caption{\label{figureHPd3}Entropy (subtracted by $\beta_HM$) and size of the string star in the presence of a spatial circle $\mathbb{S}^1_z$ for $d<3$. The star marks the critical point where the Gregory-Laflamme instability occurs. The uniform string star (given by HP$_d$) is at $M<M_\text{HP,GL}$, while the non-uniform string star lies above $M_\text{HP,GL}$. At $M\gg M_\text{HP,GL}$, the non-uniform solution becomes localized and closely approximates the solution in $d+1$ noncompact (spatial) dimensions, i.e. HP$_{d+1}$.}
\end{figure}

Besides, in the regime $M\gg M_\text{HP,GL}$, comparing the entropy of HP$_{d+1}$ to that of HP$_d$ reveals a significant difference. Specifically, since $d<3$, both the second term in (\ref{SHP}) for HP$_{d+1}$ and that for HP$_d$ are positive, and the ratio between them is given by
\begin{equation}
\label{ratio}
    \frac{S^{(d+1)}-\beta_HM}{S^{(d)}-\beta_HM}\sim \left(\left(\frac{G_N^{(d+1)}M}{l_s^2}\right)^{\frac{1}{4-d}}L\right)^{\frac{2}{3-d}}\ .
\end{equation}
For $M\gg M_\text{HP,GL}$, this ratio satisfies
\begin{equation}
    \frac{S^{(d+1)}-\beta_HM}{S^{(d)}-\beta_HM}\gg 1\ .
\end{equation}
Thus, far outside the uniform regime, the entropy of HP$_{d+1}$ exceeds that of HP$_d$. These observations support the conjecture that, at the critical point $M_\text{HP,GL}$, the uniform string star described by HP$_d$ transitions into a non-uniform string star, and becomes localized (described by HP$_{d+1}$) as $M$ further increases. In figure \ref{figureHPd3}, we sketch the entropy as well as the size of the string star solutions. Note that as calculated in the previous subsection, the phase transition at the critical point is first-order, which implies a discontinuous change in entropy at $M_\text{HP,GL}$.
\subsubsection{$3<d<4$}\label{sec3d4}
For $d$ between 3 and 4, the regime of validity for HP$_d$ is $M_g\ll M\ll M_c$, while for HP$_{d+1}$, the regime is $M_c^{(d+1)}\ll M\ll M_g^{(d+1)}$, where $M_c^{(d+1)}$ and $M_g^{(d+1)}$ are defined in (\ref{Mcgd1}). Once again, we have the inequality $M_c^{(d+1)}\ll M_g^{(d+1)}$ due to the condition $g_s^2L/l_s\ll 1$ (see (\ref{gsL})). The hierarchy of scales now becomes
\begin{equation}
\label{hi3d4}
    M_c^{(d+1)},M_g\ll M_\text{HP,GL}\ll M_c,M_g^{(d+1)}\ ,\quad 3<d<4\ ,
\end{equation}
where the relative ordering between $M_c^{(d+1)}$ and $M_g$, as well as that between $M_c$ and $M_g^{(d+1)}$, remains undetermined. Note that the inequalities $M_c^{(d+1)}\ll M_\text{HP,GL}\ll M_c$ follow from the condition $L\gg l_s$, while the inequalities among $M_g\ll M_\text{HP,GL}\ll M_g^{(d+1)}$ are derived from (\ref{GLcon}).

From (\ref{MhatC}) and (\ref{MHPGL}), it follows that for HP$_{d+1}$ to be small enough to fit within the compact dimension, the mass $M$ has to be much smaller than $M_\text{HP,GL}$. However, when $M<M_\text{HP,GL}$, the system remains in the uniform region (see (\ref{uniform})). This implies that the uniform string star described by HP$_d$ cannot transition into the localized string star as $M$ increases. This is not surprising, as (\ref{SHP}) already shows that $S(\text{HP}_d)>\beta_HM$ while $S(\text{HP}_{d+1})<\beta_HM$, indicating that the entropy of uniform string star exceeds that of the localized one. 

Then, it remains unclear what kind of solutions arises from the Gregory-Laflamme instability of the uniform string star for $3<d<4$. One possibility is that the uniform string star transitions into a black hole or black string at the critical point. In fact, using (\ref{MBSGL}), (\ref{MHPGL}) and the condition $L\gg l_s$, one can derive that
\begin{equation}
    M_\text{HP,GL}\ll M_\text{BS,GL}\ .
\end{equation}
Therefore, at $M_\text{HP,GL}$, a localized black hole exists and can be well approximated by a $(d+2)$-dimensional Schwarzschild black hole. Note that string corrections to the black hole remain small, as $M_c^{(d+1)}\ll M_\text{HP,GL}$.

We can further compare the entropy of HP$_d$ with that of the black hole at the critical point $M_\text{HP,GL}$. Using (\ref{SHP}), the entropy of HP$_d$ scales as $L^{d-4}l_s^3/G_N^{(d+1)}$. In contrast, from (\ref{SBH}) (with $d$ replaced by $d+1$), the black hole entropy is of order $L^{\frac{d^2-4d+1}{d-1}}l_s^{\frac{2d}{d-1}}/G_N^{(d+1)}$.\footnote{I thank Roberto Emparan, Mikel Sanchez-Garitaonandia and Marija Toma\v{s}evi\'{c} for pointing out an error in the calculation in version v1 of this paper.} Given $L\gg l_s$, the black hole solution exhibits much higher entropy, making it feasible for the uniform string star to transition into a black hole. Furthermore, the disparity between these two entropies at the critical point suggests that the uniform string star undergoes a first-order phase transition. Notably, this conclusion is consistent with the findings in the previous subsection.
\begin{figure}
	\centering
	\subfigure[]{
	\begin{minipage}[t]{0.45\linewidth}
	\centering
	\includegraphics[width=2.5in]{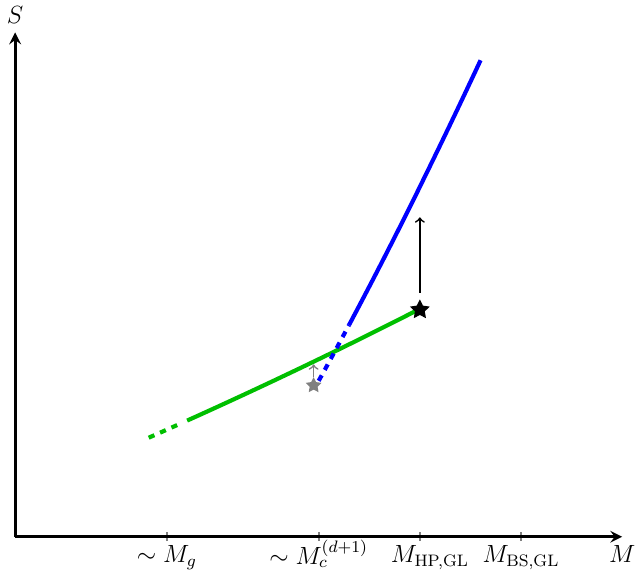}\label{entropiesHP3d4}
	\end{minipage}}
	\subfigure[]{
	\begin{minipage}[t]{0.45\linewidth}
	\centering
	\includegraphics[width=2.5in]{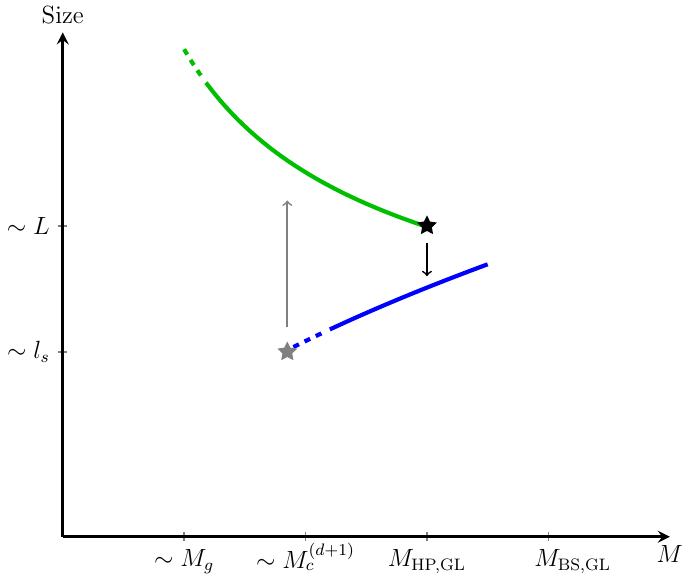}\label{sizeHP3d4}
	\end{minipage}}
	\centering
\caption{\label{figureHP3d4}Entropies and sizes of the uniform string star and the localized black hole in $\mathbb{R}^{d,1}\times\mathbb{S}^1_z$ for $3<d<4$, represented by the green line and the blue line, respectively. The black star marks the critical point where the uniform string star encounters a Gregory-Laflamme instability and transitions into the black hole. The gray star denotes a conjectured critical point at which the black hole encounters an instability and transitions into the uniform string star.}
\end{figure}

In summary, as we sketch in figure \ref{figureHP3d4}, we propose the following scenario for $3<d<4$, analogous to hysteresis: as the mass increases, the uniform string star (HP$_d$) transitions into a higher-dimensional Schwarzschild black hole at the critical mass $M_\text{HP,GL}$. And starting with a $(d+2)$-dimensional Schwarzschild black hole and decreasing $M$ would lead to an instability at another critical mass below $ M_\text{HP,GL}$, presumably around $M_c^{(d+1)}$. The outcome of this instability depends on the relative ordering of the scales $M_c^{(d+1)}$ and $M_g$. Specifically, if $M_c^{(d+1)}\gg M_g$, the black hole would transition to the uniform string star, while for $M_c^{(d+1)}\ll M_g$, it would become a free string. The scenario depicted in figure \ref{figureHP3d4} corresponds to the former case.
\subsubsection{$4<d<5$}
\begin{figure}
	\centering
	\subfigure[]{
	\begin{minipage}[t]{0.45\linewidth}
	\centering
	\includegraphics[width=2.5in]{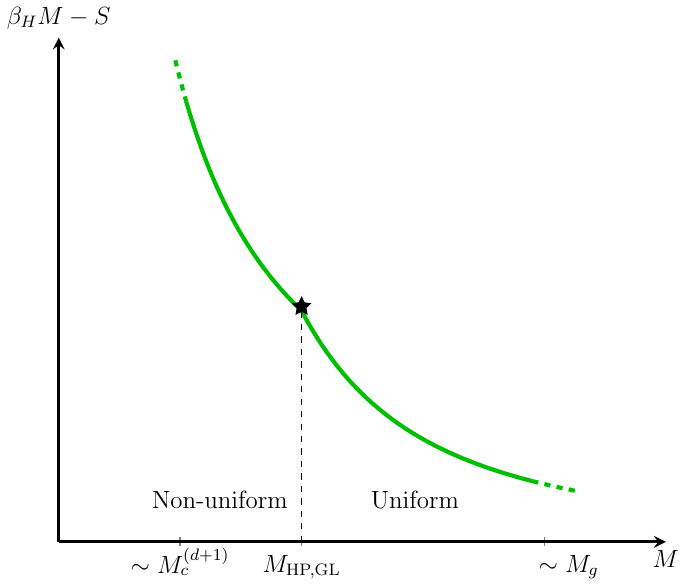}\label{entropiesHP4d5}
	\end{minipage}}
	\subfigure[]{
	\begin{minipage}[t]{0.45\linewidth}
	\centering
	\includegraphics[width=2.5in]{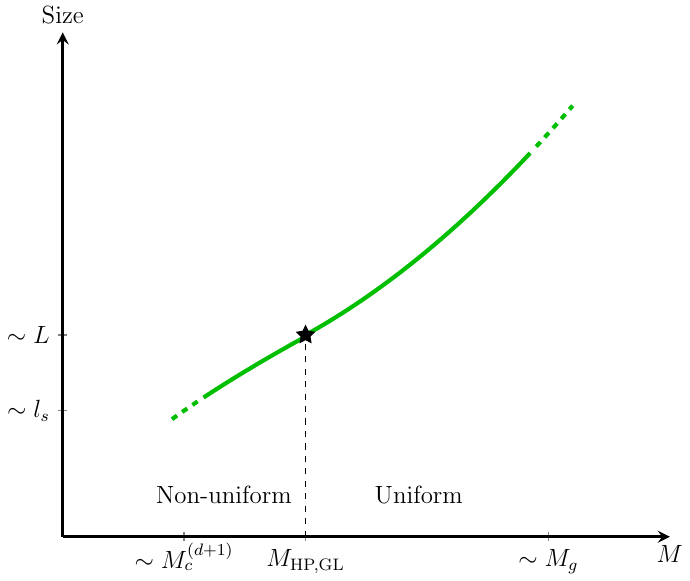}\label{sizeHP4d5}
	\end{minipage}}
	\centering
\caption{\label{figureHP4d5}$\beta_HM-S$ and size as functions of $M$ for the string star in $\mathbb{R}^{d,1}\times\mathbb{S}^1_z$ at $4<d<5$. Above the critical mass $M_\text{HP,GL}$, the solid green line corresponds to the uniform string star solution (HP$_d$), while below $M_\text{HP,GL}$, it represents the non-uniform string star solution. For $M\ll M_\text{HP,GL}$, the non-uniform solution behaves as the string star in $(d+2)$-dimensional noncompact spacetime (HP$_{d+1}$).}
\end{figure}
In this case, the regimes of validity for HP$_d$ and HP$_{d+1}$ are respectively $M_c\ll M\ll M_g$ and $M_c^{(d+1)}\ll M\ll M_g^{(d+1)}$. The hierarchy of scales is as follows.
\begin{equation}
    M_c^{(d+1)}\ll M_c\ll M_\text{HP,GL}\ll M_g\ll M_g^{(d+1)}\ ,\quad 4<d<5\ .
\end{equation}

When $M\ll M_\text{HP,GL}$, which is outside the uniform region (\ref{uniform}), the size of HP$_{d+1}$ is much smaller than $L$, allowing it to fit well within the compact dimension. In this regime, we can compare the entropies of HP$_d$ and HP$_{d+1}$. First, the second term in (\ref{SHP}) is negative for both HP$_d$ and HP$_{d+1}$. Furthermore, the ratio (\ref{ratio}) is much smaller than 1 for $M\ll M_\text{HP,GL}$. Therefore, HP$_{d+1}$ has a higher entropy than HP$_d$. This suggests that the uniform string star does transition into a localized string star (HP$_{d+1}$) as $M$ decreases. To conclude, we sketch the entropy as well as the size with respect to the mass as in figure \ref{figureHP4d5}. Note that it is calculated in the previous subsection that the phase transition at the critical point is second-order, implying that the entropy changes continuously.
\subsubsection{$d$ close to 5}
For $d$ near 5, as discussed in section \ref{secd6}, the solution HP$_{d+1}$ requires modifications. In particular, from (\ref{d6HP}), the length scale of the modified HP$_{d+1}$ is given by $l_s/\sqrt{\chi(0)}$. Outside the uniform region (\ref{uniform}), particularly when $M\ll M_\text{HP,GL}$, using (\ref{Mchi0}) we find that $\chi(0)\gg \alpha'/L^2$. Therefore, in this regime the size of the modified HP$_{d+1}$ is much smaller than $L$, which means that a localized string star exists.

\begin{figure}
	\centering
	\subfigure[]{
	\begin{minipage}[t]{0.45\linewidth}
	\centering
	\includegraphics[width=2.5in]{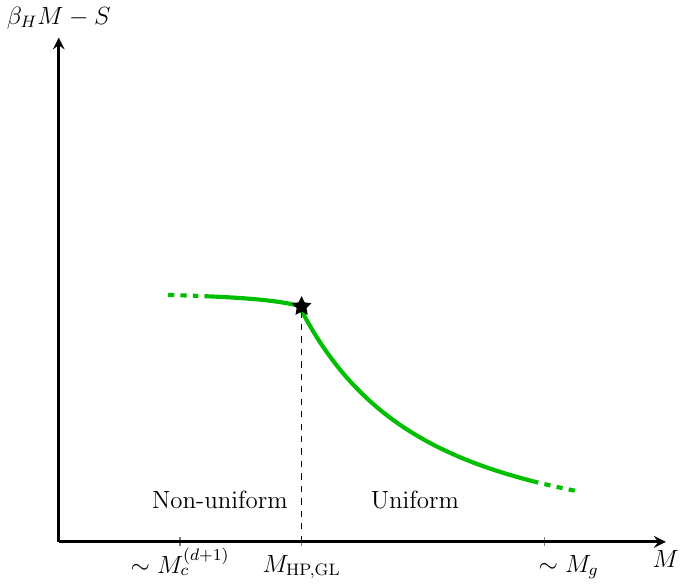}\label{entropiesHPd5}
	\end{minipage}}
	\subfigure[]{
	\begin{minipage}[t]{0.45\linewidth}
	\centering
	\includegraphics[width=2.5in]{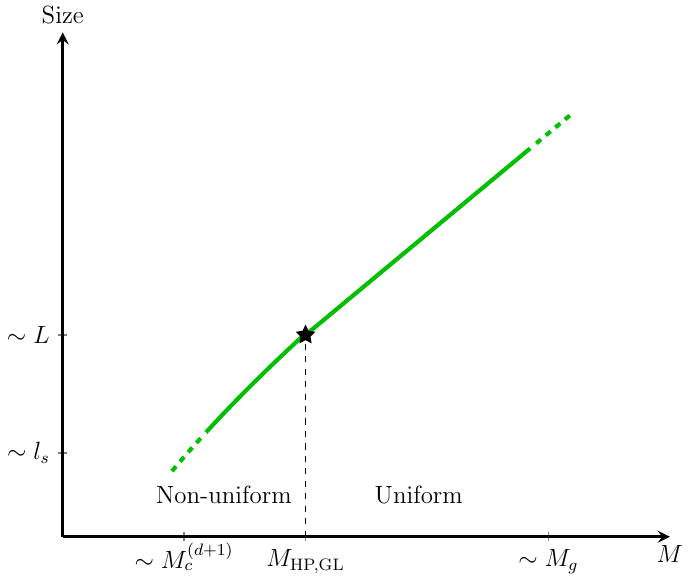}\label{sizeHPd5}
	\end{minipage}}
	\centering
\caption{\label{figureHPd5}$\beta_HM-S$ and size of the string star in $\mathbb{R}^{d,1}\times\mathbb{S}^1_z$ for $d$ close to 5. The uniform solution exists above $M_\text{HP,GL}$, while it transitions to a non-uniform string star below this critical mass. This non-uniform solution becomes localized in the spatial circle in the regime $M\ll M_\text{HP,GL}$.}
\end{figure}
The entropies of HP$_d$ and (modified) HP$_{d+1}$ are given by (\ref{SHP}) and (\ref{SHPl6}), respectively. It can be seen that both $S(\text{HP}_{d+1})-\beta_HM$ and $S(\text{HP}_d)-\beta_HM$ are negative, and the ratio between them
\begin{equation}
    \frac{S^{(d+1)}-\beta_H}{S^{(d)}-\beta_H}\sim \frac{G_N^{(6)}M}{L\alpha'}\ ,
\end{equation}
where we take $d \to 5$. This ratio is much smaller than 1 when the mass $M$ is much less than $M_\text{HP,GL}\sim\alpha'L/G_N^{(6)}$. Therefore, $S(\text{HP}_{d+1})>S(\text{HP}_d)$ and HP$_{d+1}$ dominates. This suggests that at the critical point $M_\text{HP,GL}$, the uniform string star transitions into a non-uniform string star, which then becomes localized as $M$ continues to decrease. This scenario is illustrated in figure \ref{figureHPd5}. Note that the entropy is continuous at the critical point $M_\text{HP,GL}$, since the phase transition is second-order as calculated in the previous subsection.

\section{A more complete picture of the transition process}\label{seccomplete}
Now, combining the discussions from the previous two sections, we can present a comprehensive picture of the transitions that occur as the mass of a large uniform black string decreases, under the compact dimension length constraints $L\gg l_s$ and (\ref{GLcon}). First, for the case where $2<d<3$, we merge the results shown in figures \ref{entropiesBS} and \ref{entropiesHPd3}, and add the free string. The resulting picture is depicted in figure \ref{entropiesd3}, where, as the mass of the black string decreases, it first encounters a Gregory-Laflamme instability at $M_\text{BS,GL}$, becoming either a non-uniform black string or a localized black hole. The system then transitions into a non-uniform string star around $M\sim M_c^{(d+1)}$. As the mass decreases further, the solution reverts to a uniform configuration near $M_\text{HP,GL}$ (since the phase transition from uniform to non-uniform string star is first-order, the reverse transition does not occur exactly at $M_\text{HP,GL}$). Finally, the uniform string star evolves into a free string. Therefore, the sequence of transitions leading up to the free string is uniform-non-uniform-uniform, as indicated in figure \ref{entropiesd3}.
\begin{figure}
	\centering
\includegraphics[scale=0.8]{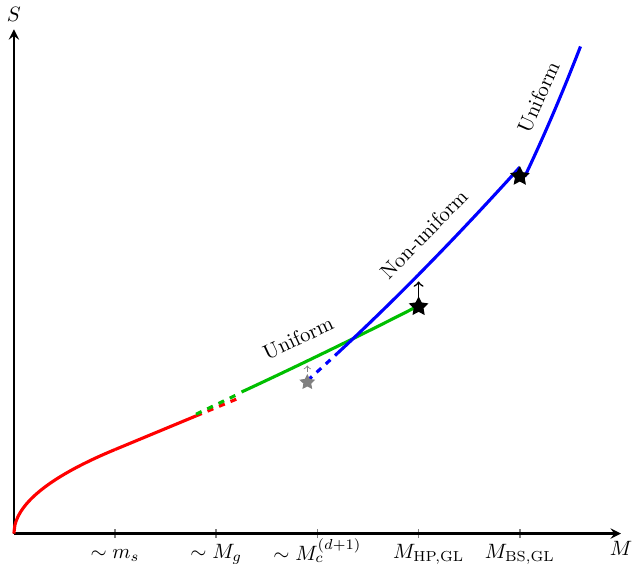}
\caption{\label{entropies3d4}Entropies of black objects (blue lines), the uniform Horowitz-Polchinski solution (green lines) and the free string (red line) in $\mathbb{R}^d\times\mathbb{S}^1_z$ for $3<d<4$. The black stars mark the critical points where the Gregory-Laflamme instabilities occur. The gray star denotes the conjectured critical point at which the black hole encounters an instability and transitions into the uniform Horowitz-Polchinski solution (or the free string).}
\end{figure}

For $3<d<4$, by incorporating the string star as depicted in figure \ref{entropiesHP3d4}, along with the free string into figure \ref{entropiesBS}, we can sketch the entropies of the black objects, string stars and free string, as shown in figure \ref{entropies3d4}. As before, at $M_\text{BS,GL}$, the black string encounters a Gregory-Laflamme instability, transitioning into a non-uniform black string or a black hole localized in the $z$ dimension. Subsequently, as discussed in section \ref{sec3d4}, we conjecture that at a mass of order $M_c^{(d+1)}$, the localized black hole encounters an instability, evolving into a uniform string star. Here, we have assumed that $M_c^{(d+1)}\gg M_g$. If this assumption is relaxed, it is also possible for the black hole to transition directly into a free string. In such a scenario, the sequence of phase transitions would follow a simpler uniform-to-non-uniform pattern.

\begin{figure}
	\centering
\includegraphics[scale=0.8]{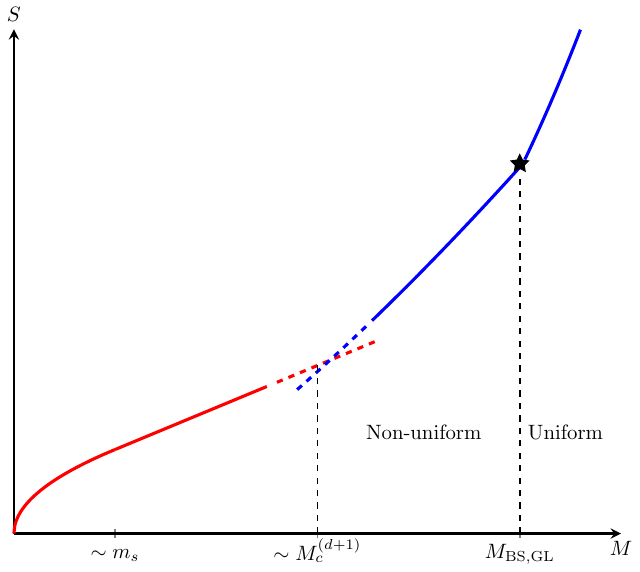}
\caption{\label{entropiesBSHPdg4}Entropies of black objects (blue lines) and the free string (red line) in $\mathbb{R}^d\times\mathbb{S}^1_z$ for $d\ge 4$. The black star marks the critical point where the Gregory-Laflamme instability occurs.}
\end{figure}
The scenario is much simpler for the case of $d>4$. As discussed in section \ref{secfree} (see also sections \ref{sechp} and \ref{secd6}), the self-gravitating string does not participate in the transition between the black hole and the free string. This results in the phase transition pattern illustrated in figure \ref{entropiesBSHPdg4}, where, as the mass decreases, the uniform black string first transitions into a non-uniform black object at $M_\text{BS,GL}$ and subsequently evolves into the free string in the regime where $M\sim M_c^{(d+1)}$.

\subsection{$d=3$ and $4$}
Previously, we treated $d$ as a continuous parameter and separately discussed the cases of $2<d<3$ (figure \ref{entropiesd3}), $3<d<4$ (figure \ref{entropies3d4}) and $d>4$ (figure \ref{entropiesBSHPdg4}). However, these discussions excluded two integer values: $d=3$ and $d=4$. In this subsection, we will address these specific cases.

We begin by considering the $d=3$ case. As discussed in section \ref{secons}, for $2<d<3$, $M_\text{HP,GL}$ is much smaller than $M_c^{(d+1)}$ (see (\ref{hi2d3})), while for $3<d<4$, it is much larger than $M_c^{(d+1)}$ (see (\ref{hi3d4})). This implies that \textit{at} $d=3$, $M_\text{HP,GL}$ is of the same order as $M_c^{(d+1)}$. Indeed, from (\ref{MHPGL}) and (\ref{Mcgd1}),
\begin{equation}
    M_\text{HP,GL}\sim M_c^{(d+1)}\sim M_g^{(d+1)}\sim \frac{1}{g_s^2L}\ ,\quad d=3\ .
\end{equation}

Besides, for $2<d<3$, the non-uniform string star serves as an intermediate phase between the black hole and the uniform string star as in figure \ref{entropiesd3}, whereas for $3<d<4$ it does not play a role in the transition between the black hole and the uniform string star (see figure \ref{entropies3d4}). This implies that as $d\to 3^-$, the mass range of the non-uniform string star gradually shrinks. And at $d=3$, this range collapses to the regime $M\sim M_c^{(d+1)}$, which aligns with the fact that the mass of the four-dimensional Horowitz-Polchinski solution remains constant, as inferred from (\ref{MhatC}). 
\begin{figure}
	\centering
\includegraphics[scale=0.8]{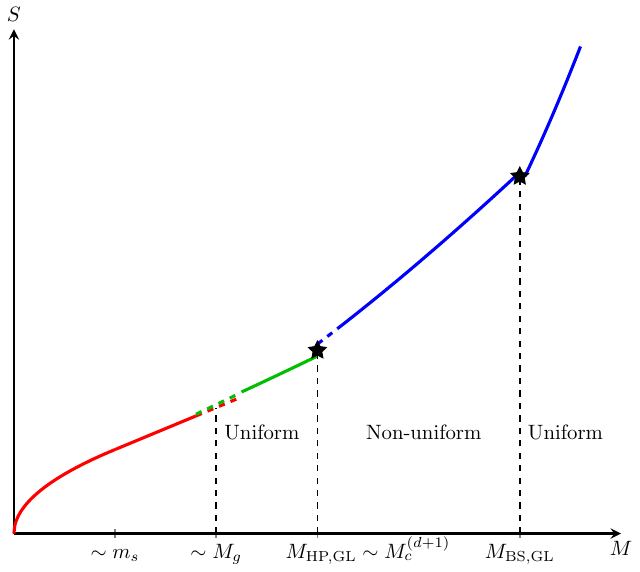}
\caption{\label{entropiesde3}Entropies of black objects (blue lines), the uniform string star (green line) and the free string (red line) in $\mathbb{R}^d\times\mathbb{S}^1_z$ for $d=3$. The black stars mark the critical points where the Gregory-Laflamme instabilities occur.}
\end{figure}

From the observations above, we can conclude that at $d=3$, a transition between the localized black hole and the uniform string star occurs at the mass scale $M_c^{(d+1)}$. To summarize the complete transition sequence, in the $d=3$ case, the uniform black string undergoes a series of phase transitions-uniform to non-uniform to uniform-as mass decreases, as sketched in figure \ref{entropiesde3}. Specifically, it first transitions into a non-uniform black string or a black hole, then into the uniform string star, and ultimately becomes the free string.

Now, we discuss the case where $d=4$. First, as $d$ approaches 4 from below, the upper bound of the uniform string star, $M_g$, converges towards the critical point associated with the Gregory-Laflamme instability, namely $M_\text{HP,GL}$. In fact, as mentioned earlier, the four-dimensional Horowitz-Polchinski solution has a constant mass, $M\sim 1/g_s^2l_s$. Furthermore, given $L\gg l_s$, we have $ M_c^{(d+1)}\sim 1/g_s^2L\ll 1/g_s^2l_s$ for $d=4$. Thus, as the mass decreases, the black hole directly transitions into the free string at $M\sim M_c^{(d+1)}$, following a pattern of phase transitions similar to those illustrated in figure \ref{entropiesBSHPdg4} and consistent with the analysis for $d>4$. Nevertheless, if we start with the free string and increase the mass, it likely does not immediately transition into the black hole in the regime $M_c^{(d+1)}$. Instead, a black hole will be formed at around $M_\text{HP,GL}\sim M_c\sim M_g$.
\subsection{Large $d$}
So far, we have assumed that $d$ is a parameter of order 1. For completeness, we now turn to the case of large $d$.

First, in the absence of the extra spatial circle $\mathbb{S}^1_z$, the large $d$ black hole can be studied by reducing it on the $(d-1)$-sphere. This yields a two-dimensional black hole that can be analyzed exactly in $\alpha'$ using the $SL(2)_k/U(1)$ WZW model~\cite{Emparan:2013xia,Chen:2021emg,Halder:2024gwe}. Specifically, we can define a new radial coordinate, $\rho$, by
\begin{equation}
\label{rho}
    \cosh^2\rho=\left(\frac{r}{r_H}\right)^d\ .
\end{equation}
This allows us to rewrite the Schwarzschild metric (\ref{schw}) as
\begin{equation}
\label{schwd}
    ds^2=-\tanh^2\rho dt^2+\left(\frac{2r_H}{d}\right)^2d\rho^2+r_H^2\cosh^{\frac{4}{d}}\rho d\Omega_{d-1}^2\ ,
\end{equation}
where only the leading order terms in $1/d$ has been retained. This approximation is valid in the near-horizon region, where $\rho\ll d$, or equivalently $r/r_H-1\ll 1$.

By further reducing on the sphere part in (\ref{schwd}), we obtain a two-dimensional dilaton field $\phi_2$ as follows.
\begin{equation}
\label{dil}
    e^{-2\phi_2}\propto\omega_{d-1}r^{d-1}\sim \cosh^2\rho\ .
\end{equation}

By parametrizing $r_H$ as
\begin{equation}
\label{rHk}
    r_H=\frac{\sqrt{k\alpha'}}{2}d
\end{equation}
the two-dimensional metric is
\begin{equation}
    ds^2=-\tanh^2\rho dt^2+k\alpha'd\rho^2\ .
\end{equation}
This metric, along with the dilaton (\ref{dil}), corresponds precisely to that of the two-dimensional black hole described by the $SL(2)_k/U(1)$ model. Note that the metric is valid in the regime where $k\gg 1$, or equivalently, $r_H/l_s\gg d$ as can be seen from (\ref{rHk}).

For finite $k$, one can continue analyzing the near-horizon region using the $SL(2)_k/U(1)$ model. In this model, the Hawking temperature of the black hole is given by
\begin{equation}
    \beta=2\pi \sqrt{k}l_s\ .
\end{equation}
When $k$ approaches 3 for bosonic strings or 1 for type II strings, the winding tachyon becomes not normalizable~\cite{Chen:2021emg}. Notably, it has been proposed that when $k$ decreases to these values, the black hole transitions into highly excited strings~\cite{Giveon:2005mi} (see also~\cite{Balthazar:2021xeh}).

We now add the extra spatial circle $\mathbb{S}^1_z$. In this case, the uniform black string at large $d$ can be described by a direct product of the $SL(2)_k/U(1)$ model and a $\mathbb{S}^1_z$ sigma model, obtained by reducing on the $(d-1)$-sphere. Presumably, a Gregory-Laflamme instability may also be present, analogous to the case for finite $d$. We now investigate this possibility.

For convenience, we Wick rotate $\tau=it$ and proceed with the analysis using the Euclidean $SL(2)_k/U(1)$ model. We focus on the bosonic string case. Similar as before, to demonstrate a Gregory-Laflamme instability, we need to find a time-independent mode with a wavenumber $k_z=2\pi/L$. In the worldsheet theory, this mode can be represented by a vertex operator of the form
\begin{equation}
\label{Vjm}
    V_{j,m,\bar m}e^{ik_zZ}\ ,
\end{equation}
where $V_{j,m,\bar m}$ is a normalizable vertex operator in the $SL(2)_k/U(1)$ model. The quantum numbers appearing in the vertex operator are given by
\begin{equation}
\label{vo}
\begin{split}
   & m=\frac{1}{2}(wk-n)\ ,\ \bar{m}=\frac{1}{2}(wk+n)\ ,   \\
    &j\in   (|m|-\mathbb{Z}^+)\cap (|\bar m|-\mathbb{Z}^+)\cap\left(-\frac{1}{2},\frac{k-3}{2}\right)\ ,
\end{split}
\end{equation}
where $w$ and $n$ represent the winding number and momentum number along the $\tau$ direction, respectively. For a $\tau$-independent mode, we set $n=0$. Furthermore, from (\ref{vo}), to satisfy the constraint $-1/2<j<(k-3)/2$ on $j$, the winding number $w$ must be nonzero. 

A physical state should satisfy the condition that the conformal dimension of the corresponding vertex operator is equal to one. We now examine whether (\ref{Vjm}) can meet this requirement. For the case where $n=0$, the conformal dimension is given by
\begin{equation}
\label{cdh}
   h=-\frac{j(j+1)}{k-2}+\frac{w^2k}{4}+\frac{\alpha'k_z^2}{4}\ . 
\end{equation}
Note that the r.h.s. is a monotonically decreasing function of $j$ in the range $(-1/2,(k-3)/2)$ ($k>3$). Therefore, by plugging $(k-3)/2$ in (\ref{cdh}), we can establish a lower bound for $h$ as follows.
\begin{equation}
\label{hjmax}
\begin{split}
    h&>-\frac{(k-3)(k-1)}{4(k-2)}+\frac{w^2k}{4}+\frac{\alpha'k_z^2}{4}\\
    &=\frac{1}{4}\left((w^2
    -1)k+2+\frac{1}{k-2}\right)+\frac{\alpha'k_z^2}{4}\ .
\end{split}
\end{equation}

For $w\ge 2$, the inequality (\ref{hjmax}) yields
\begin{equation}
\begin{split}
    h>\frac{1}{4}\left(3k+2+\frac{1}{k-2}\right)+\frac{\alpha'k_z^2}{4}\ge 2+\frac{\sqrt{3}}{2}+\frac{\alpha'k_z^2}{4}>1+\frac{\alpha'k_z^2}{4}\ .
\end{split}
\end{equation}
This implies that there is no real $k_z$ that can satisfy the physical condition $h=1$. For $w=1$, the maximum possible value of $j$, satisfying the constraint in (\ref{vo}), is $j_\text{max}=k/2-2$. Therefore, using the monotonicity of the conformal dimension in (\ref{cdh}), we find
\begin{equation}
    h\ge -\frac{j_\text{max}(j_\text{max}+1)}{k-2}+\frac{w^2k}{4}+\frac{\alpha'k_z^2}{4}=1+\frac{\alpha'k_z^2}{4}\ .
\end{equation}
Thus, the only physical state corresponds to $j=k/2-2$ and $k_z=0$. This implies that there exists no non-uniform mode. In conclusion, the $SL(2)_k/U(1)$ model does not exhibit a Gregory-Laflamme instability.

Finally, we note that in gravity theory, if we keep only the leading-order terms in $1/d$ in the linearized Einstein equations for non-uniform perturbation in the near-horizon region, we find only a time-independent solution at $\beta=0$~\cite{Asnin:2007rw}. This implies that no Gregory-Laflamme instability occurs in the uniform black string at large $d$, consistent with the result of our analysis in this subsection. However, when the $O(1/d)$ terms are also included in the linearized Einstein equations, which are not captured by the $SL(2)_k/U(1)$ model, a nonzero critical temperature, $\beta_\text{GL}^2=4L^2/d$, can be obtained~\cite{Kol:2004pn,Asnin:2007rw}. Nevertheless, based on the gravity result, it is reasonable to speculate that for $d\gg L^2/l_s^2$, $\beta_\text{GL}$ is well below the temperature at which the black string transitions into highly excited strings. In this case, the black string would transition into fundamental strings without experiencing the Gregory-Laflamme instability.

\section{Discussion}\label{secdis}
Our goal in this paper was to explore the transitions from large uniform black strings to fundamental strings in the presence of an extra spatial circle $\mathbb{S}^1_z$. The uniform black string, when reduced on $\mathbb{S}^1_z$, corresponds to a Scharzschild black hole. We focus on the regime where the size of the spatial circle is much larger than the string length, $L\gg l_s$. In this context, there is a rich phase structure that includes both uniform and non-uniform solutions.

Gregory-Laflamme demonstrated that the uniform black string is unstable towards non-uniformity when its mass is smaller than a critical value, $M_\text{BS,GL}$. Similarly, we showed that the uniform Horowitz-Polchinski solution in $\mathbb{R}^d\times\mathbb{S}^1_z$ exhibits thermodynamic instability towards non-uniformity. In Lorentzian signature, this can be interpreted as the Gregory-Laflamme instability of a uniform string star.

We performed a perturbative analysis of the non-uniform Horowitz-Polchinski solution. From the leading-order perturbations, $\delta\chi^{(1)}(r,z)$ and $\delta\varphi^{(1)}(r,z)$, to the uniform solutions $\chi_*(r)$ and $\varphi_*(r)$, we determined the critical mass $M_\text{HP,GL}$ of the Gregory-Laflamme instability (section \ref{secHPLO}). This critical value provides a boundary for the stable region of the uniform solution, which we define as the uniform region. For $2<d<4$, the uniform region is bounded from above, i.e., $M<M_\text{HP,GL}$, while for $4<d<6$, it is bounded from below, i.e., $M>M_\text{HP,GL}$. Proceeding to the next-to-leading-order non-uniform perturbations, $\delta\chi^{(2)}(r,z)$ and $\delta\varphi^{(2)}(r,z)$, we determined the order of the phase transition at the critical mass $M_\text{HP,GL}$ (section \ref{secHPNLO}). We found that for $2<d<4$, the phase transition is first-order, whereas for $4<d<6$, it is second-order.

We also discussed possible non-uniform solutions arising from the uniform string star at the critical point of the Gregory-Laflamme instability (section \ref{secons}). For $2<d<3$, the uniform solution transitions into a non-uniform string star. This non-uniform solution is disconnected from the uniform one, since the phase transition is first-order. Therefore, the perturbative analysis is insufficient to obtain the resulting non-uniform solution, and one needs to solve the full equations of motion from the Horowitz-Polchinski effective action (\ref{Sphichiz}) in $\mathbb{R}^d\times\mathbb{S}^1_z$. We leave this to future work.\footnote{After this paper was submitted to arXiv, a new paper appeared, which finds the non-uniform Horowitz-Polchinski solutions beyond the perturbative analysis~\cite{Emparan:2024mbp}.}  Nevertheless, as the mass continues to increase and moves away from $M_\text{HP,GL}$, the non-uniform string star becomes localized in the compact dimension and can be well approximated by the Horowitz-Polchinski solution in $\mathbb{R}^{d+1}$, denoted by HP$_{d+1}$ in this paper.

For $3<d<4$, however, we showed that the uniform string star cannot transition into a non-uniform string star. Instead, it is likely to transition into a black hole localized in $\mathbb{S}^1_z$ at the critical point $M_\text{HP,GL}$. We verified that the localized black hole has a much higher entropy, consistent with the earlier finding that the phase transition at $M_\text{HP,GL}$ is first-order. Notably, this scenario establishes a direct connection between the black hole and the string star, in a regime where both solutions are valid. This contrasts with the situation in noncompact dimensions, where neither solution is valid in the transition regime $M\sim M_c$.

For $4<d<6$, the uniform string star transitions continuously into a non-uniform string star as the mass decreases to $M_\text{HP,GL}$, since the phase transition is second-order. The resulting non-uniform solution near the critical mass can be analyzed perturbatively. If $d<5$, the non-uniform solution can be well approximated by the higher dimensional string star described by HP$_{d+1}$, for $M\ll M_\text{HP,GL}$. When $d$ is close to 5, for solutions with $M\ll M_\text{HP,GL}$, it becomes necessary to include the quartic terms in the effective action, similar to~\cite{Balthazar:2022hno}. For $d$ strictly larger than 5, however, the effective action breaks down when $M$ decreases away from $M_\text{HP,GL}$.

By piecing together various black objects and string stars, we obtained a more complete picture of transition from the uniform black string to fundamental strings (section \ref{seccomplete}). Specifically, starting with a large uniform black string and decreasing its mass, it first transitions into a non-uniform black string or a black hole at $M_\text{BS,GL}$. For $2<d<3$, as shown in figure \ref{entropiesd3}, the non-uniform black object transitions into a non-uniform string star at the scale $M_c^{(d+1)}$, then becomes a uniform string star near $M_\text{HP,GL}$, and eventually evolves into free strings at the scale $M_g$. For $d=3$, the regime for the non-uniform string star shrinks to the scale $M_c^{(d+1)}$, allowing the non-uniform black object to transition directly into a uniform string star at this scale, as shown in figure \ref{entropiesde3}. A similar transition scenario holds for $3<d<4$ with $M_g\ll M_c^{(d+1)}$ (see figure \ref{entropies3d4}). For $d\ge 4$ or $3<d<4$ with $M_g\gg M_c^{(d+1)}$, fundamental strings are free around $M_c^{(d+1)}$, and thus the non-uniform black object directly transitions into free strings at this scale, as illustrated in figure \ref{entropiesBSHPdg4}.

We also analyzed the uniform black string at large $d$ using the $SL(2)_k/U(1)$ WZW model. Our result showed that the vertex operators corresponding to non-uniform perturbations fail to meet the physical condition that their scaling dimensions equal 1. Note that this analysis remains valid even when $L$ is not much larger than $l_s$. Furthermore, since the $SL(2)_k/U(1)$ model captures only the leading-order physics (in $1/d$) in the near-horizon region, we cautiously concluded that the uniform black string does not undergo the Gregory-Laflamme instability before transitioning into fundamental strings, provided that $d\gg L^2/l_s^2$.

Apart from the black string solutions, another solution asymptotic to $\mathbb{R}^{d,1}\times\mathbb{S}^1$ is the static Kaluza-Klein bubble~\cite{Elvang:2004iz}. In appendix \ref{sec2HP}, we turn on temperature and study string stars in this background. We find that solutions exist for any $d$, and are valid as long as $L\gg l_s$. It is interesting to note that these solutions break the winding symmetries along both the $\tau$ and $z$ circles. Similar solutions exhibiting double winding symmetry breaking exist in thermal AdS$_3$, as explored in~\cite{Urbach:2023npi}. Furthermore, a worldsheet theory is conjectured in~\cite{Halder:2023nlp} to describe this kind of solutions in thermal AdS$_3$. It is intriguing to further investigate the relationship between these two kinds of solutions.

In this paper, for finite $d$, we focus on a specific regime $l_s\ll L\ll l_s/g_s^{2/(6-d)}$, with $g_s\ll 1$ for weakly-coupled string theories. When the upper bound is not satisfied, quantum effects of the Horowitz-Polchinski solution must be considered to obtain the critical mass $M_\text{HP,GL}$. On the other hand, if the lower bound is not satisfied, e.g., $L\sim l_s$, a worldsheet analysis is necessary, similar to the treatment for large $d$ in the uniform black string scenario. 

When the string coupling is not small, we must study M-theory, introducing one more spatial circle, $\mathbb{S}^1_y$. The circumference of this new circle is proportional to the string coupling. In the strong string coupling regime, $\mathbb{S}^1_y$ is large, and M-theory becomes supergravity in the higher-dimensional spacetime. In this regime, the uniform black string considered in this paper is uplifted to a uniform black brane wrapping both the $y$ and $z$ circles. As the mass decreases, a large uniform black brane encounters two possible Gregory-Laflamme instabilities, each corresponding to non-uniformity along one of the two spatial circles. Which instability occurs first depends on the relative size of the circles. For a detailed analysis of the Gregory-Laflamme instability of a black brane wrapping around a square two-torus, see~\cite{Kol:2006vu} (and~\cite{Kol:2004pn} for generalizations to arbitrary compactification manifolds).

As a final remark, it has been shown that the uniform black string, non-uniform black string and localized black hole, reviewed in section \ref{secGL}, can be dual to various phases in a finite-temperature gauge theory compactified on a spatial circle~\cite{Aharony:2004ig,Harmark:2004ws}. It would be intriguing to investigate whether—and how—the uniform and non-uniform string stars discussed in this paper manifest in the dual gauge theory. This could open new avenues for understanding the full phase structure from the gauge theory perspective. We leave this exploration to future work.
\section*{Acknowledgement}
I would like to express my gratitude to David Kutasov for invaluable discussions and insightful comments on the draft. I thank Bruno Balthazar, Yiming Chen, Savan Kharel and Emil Martinec for helpful discussions and suggestions. I also thank Cindy Zhou for comments on the figures in this paper. This work was supported in part by DOE grant DE-SC0009924.

\appendix

\section{String stars in the thermal Kaluza-Klein bubble}\label{sec2HP}
In the presence of an extra spatial circle, in addition to the simple Kaluza-Klein spacetime $\mathbb{R}^{d,1}\times \mathbb{S}^1_z$, there exists another solution in gravity theory known as the static Kaluza-Klein bubble~\cite{Elvang:2004iz}. This solution asymptotes to $\mathbb{R}^{d,1}\times \mathbb{S}^1_z$ at infinity. Specifically, the metric takes the form
\begin{equation}
\label{tbb}
    ds^2=-dt^2+\left(1-\left(\frac{r_H}{r}\right)^{d-2}\right)dz^2+\frac{dr^2}{1-\left(\frac{r_H}{r}\right)^{d-2}}+r^2d\Omega_{d-1}^2\ ,
\end{equation}
where $r_H$ is the radius at which the $z$ circle shrinks to zero size.\footnote{Fermions obey anti-periodic boundary conditions on the $z$ circle in this spacetime.} The region inside it, i.e. $r<r_H$, does not contain spacetime. For the geometry to be smooth at $r_H$, this radius must be related to the asymptotic circumference of the $z$ circle by
\begin{equation}
\label{rHL}
    r_H=\frac{(d-2)L}{4\pi}\ .
\end{equation}

Note that the Kaluza-Klein bubble solution (\ref{tbb}) can be obtained by a double Wick rotation of the uniform black string (\ref{bsmet}), namely $t\leftrightarrow iz$. Then, similar to the discussion in section \ref{secBH}, the gravity result of the bubble solution is valid when $r_H\gg l_s$, ensuring that the curvature is small throughout the region ($r\ge r_H$), thereby suppressing quantum gravity effects and string corrections. From (\ref{rHL}), this condition yields $L\gg l_s$. Also note that the Kaluza-Klein bubble (\ref{tbb}) resembles Witten's bubble of nothing~\cite{Witten:1981gj}, which can be derived from the Schwarzschild black hole (\ref{schw}) through a different double Wick rotation.\footnote{Specifically, the metric of Witten's bubble of nothing is given by
\begin{equation}
    ds^2=\left(1-\left(\frac{r_H}{r}\right)^{d-2}\right)dz^2+\frac{dr^2}{1-\left(\frac{r_H}{r}\right)^{d-2}}+r^2ds_{\text{dS}_{d-1}}^2\ ,
\end{equation} where $ds_{\text{dS}_{d-1}}^2$ represents the metric of $(d-1)$-dimensional de Sitter spacetime with unit radius.}
However, unlike the static Kaluza-Klein bubble, the bubble of nothing expands over time.

Now, we introduce a finite temperature to the bubble solution (\ref{tbb}). This amounts to compactify the Euclidean time $\tau$, with the period equal to the inverse temperature $\beta$. In this setting, the on-shell Euclidean action is equivalent to that of the Euclidean uniform black string, with the roles of $L$ and $\beta$ exchanged. The on-shell action for the uniform black string is given by
\begin{equation}
\label{IBH}
    I_\text{BS}=\frac{\omega_{d-1}L}{16\pi G_N^{(d+2)}}\left(\frac{d-2}{4\pi}\right)^{d-2}\beta^{d-1}\ .
\end{equation}
It can be checked that by applying the relation $M=\partial_\beta I_\text{BS}$, together with $r_H=(d-2)\beta/4\pi$ and $G_N^{(d+2)}=G_N^{(d+1)}L$, one can recover the mass of the Schwarzschild black hole (\ref{M}). 

Using (\ref{IBH}) with the exchange of $\beta$ and $L$, we find that the mass of the thermal bubble is
\begin{equation}
    M_\text{KKb}=\partial_\beta I_\text{KKb}=\partial_\beta I_\text{BS}(\beta\leftrightarrow L)=\frac{\omega_{d-1}}{16\pi G_N^{(d+2)}}\left(\frac{d-2}{4\pi}\right)^{d-2}L^{d-1}\ ,
\end{equation}
Note that this is of the same order as the critical mass for the Gregory-Laflamme instability of the black string $M_\text{BS,GL}$, given in (\ref{MBSGL}). Furthermore, the entropy, calculated using the formula $S=\beta M-I$, is zero. As a result, in the microcanonical ensemble, the bubble solution never dominates over other solutions, such as the black string.

In this appendix, we present the string star solutions in the background of the thermal Kaluza-Klein bubble. For temperatures $\beta$ near $\beta_H$, the effective action on the spacetime background (\ref{tbb}) takes the same form as in (\ref{Sphichiz}). We focus on the spherically symmetric and $z$-independent solutions. In this case, we integrate over $z$ and the $d$-dimensional effective action reads
\begin{equation}
	\label{Sphichibb}
	\begin{split}
 	I_d=\frac{\beta \omega_{d-1}}{16\pi G_N^{(d+1)}}\int dr\ r^{d-1}\left[g^{rr}(\partial_r\varphi)^2+g^{rr}(\partial_r \chi)^2+(m_{\infty}^2+\frac{\kappa}{\alpha'}\varphi)\chi^2\right]\ ,
	\end{split}
	\end{equation}
where $g^{rr}=1/g_{rr}$ can be read off from (\ref{tbb}), namely
\begin{equation}
    g^{rr}=1-\left(\frac{r_H}{r}\right)^{d-2}\ .
\end{equation}

From the action (\ref{Sphichibb}), the e.o.m. are given by
        \begin{equation}
        \label{eomHPbb}
	\begin{split}
 	\left(r^{d-1}g^{rr}\chi'(r) \right)'&=r^{d-1}\left(m_{\infty}^2+\frac{\kappa}{\alpha'}\varphi(r)\right)\chi(r) \ ,\\
 	\left(r^{d-1}g^{rr}\varphi'(r) \right)' &=r^{d-1}\frac{\kappa}{2\alpha'}\chi(r)^2\ .
	\end{split}
	\end{equation}
To simplify these equations, we introduce the following rescaling (different from (\ref{resHP})):
\begin{equation}
\label{resHPbb}
	\begin{split}
	r&=\tilde{r}r_H\ ,\\
 	\chi (r)&=\frac{\sqrt{2}\alpha' }{\kappa r_H^2}\tilde \chi(\tilde r)\ ,\\
 	\varphi (r)&=\frac{\alpha'}{\kappa r_H^2}\tilde\varphi(\tilde r)\ ,
	\end{split}
	\end{equation}
which transforms (\ref{eomHPbb}) into the rescaled from:
        \begin{equation}
        \label{eomHPbbres}
	\begin{split}
 	\left[\tilde r^{d-1}\left(1-\tilde r^{2-d}\right)\tilde\chi'(\tilde r) \right]'&=\tilde r^{d-1}\left[ \left(m_\infty r_H\right)^2+\tilde\varphi(\tilde r)\right]\tilde\chi(\tilde r) \ ,\\
 	\left[\tilde r^{d-1}\left(1-\tilde r^{2-d}\right)\tilde\varphi'(\tilde r) \right]' &=\tilde r^{d-1}\tilde\chi(\tilde r)^2\ .
	\end{split}
	\end{equation}
Note that in terms of the rescaled variables, the boundary $r=r_H$ is transformed to $\tilde r=1$.

The differential equations (\ref{eomHPbbres}) can be numericaly solved by specifying the values of $\tilde\chi$ and $\tilde\varphi$ and their derivatives at the boundary $\tilde r=1$. The boundary derivatives can, in fact, be determined from the boundary values. To illustrate this, we perform a Taylor expansion of the fields near $\tilde r=1$,
\begin{equation}
\begin{split}
\tilde\chi(\tilde r)&=\tilde\chi(1)+\tilde\chi'(1)(\tilde r-1)+\cdots\ ,\\  \tilde\varphi(\tilde r)&=\tilde\varphi(1)+\tilde\varphi'(1)(\tilde r-1)+\cdots\ .    
\end{split}
\end{equation}
Substituting this expansion into (\ref{eomHPbbres}), we obtain the following boundary conditions
\begin{equation}
\label{devr1}
\begin{split}
    \tilde \chi'(1)&=\frac{1}{d-2}\left[(m_\infty r_H)^2+\tilde\varphi(1)\right]\tilde\chi(1)\ ,\\
    \tilde\varphi'(1)&=\frac{1}{d-2}\tilde\chi(1)^2\ .
\end{split}
\end{equation}
\begin{figure}
	\centering
	\subfigure[]{
	\begin{minipage}[t]{0.45\linewidth}
	\centering
	\includegraphics[width=2.5in]{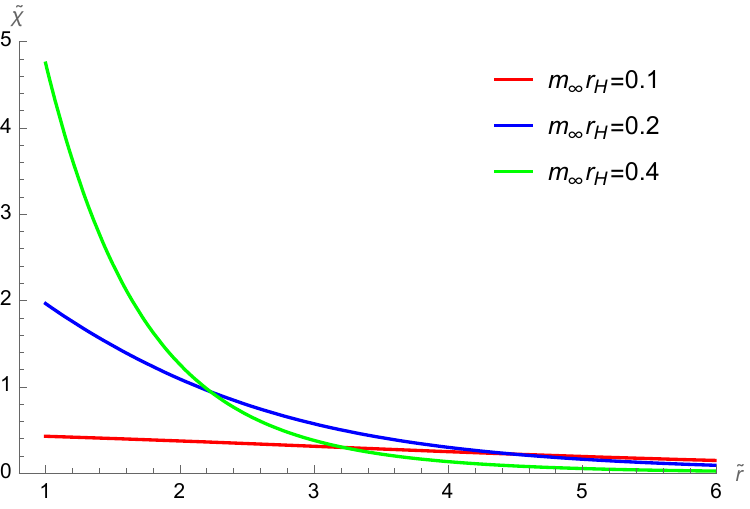}\label{tchid55}
	\end{minipage}}
	\subfigure[]{
	\begin{minipage}[t]{0.45\linewidth}
	\centering
	\includegraphics[width=2.5in]{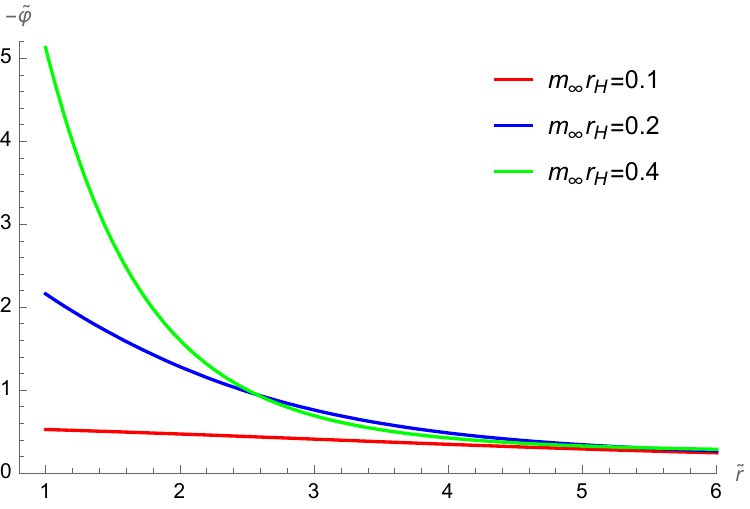}\label{tphid55}
	\end{minipage}}
	\centering
\caption{\label{tchiphid55}$\tilde\chi(\tilde r)$ and $\tilde\varphi(\tilde r)$ as functions of $\tilde r$ for $d=5.5$.}
\end{figure}
\begin{figure}
	\centering
	\subfigure[]{
	\begin{minipage}[t]{0.45\linewidth}
	\centering
	\includegraphics[width=2.5in]{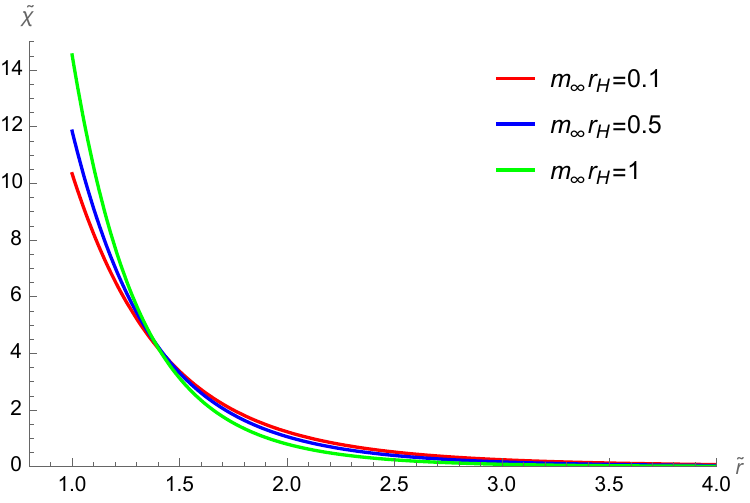}\label{tchid65}
	\end{minipage}}
	\subfigure[]{
	\begin{minipage}[t]{0.45\linewidth}
	\centering
	\includegraphics[width=2.5in]{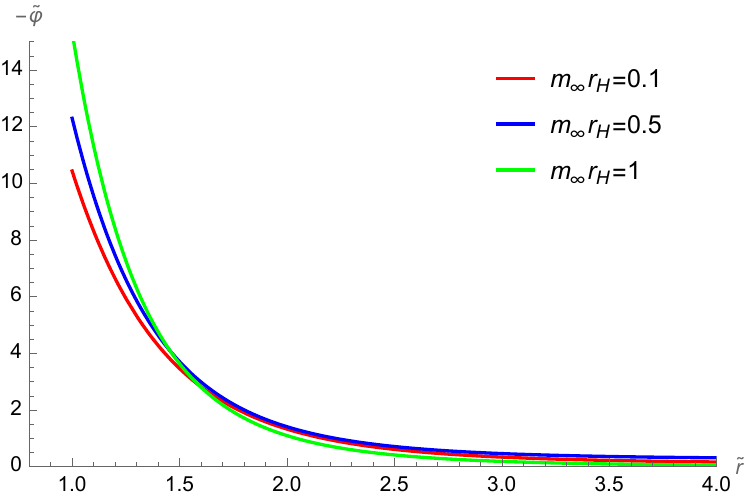}\label{tphid65}
	\end{minipage}}
	\centering
\caption{\label{tchiphid65}$\tilde\chi(\tilde r)$ and $\tilde\varphi(\tilde r)$ as functions of $\tilde r$ for $d=6.5$.}
\end{figure}

To find normalizable solutions of (\ref{eomHPbbres}), we apply the shooting method, adjusting the values of $\tilde\chi(1)$ and $\tilde\varphi(1)$ while correspondingly modifying the derivatives at $\tilde r=1$ according to (\ref{devr1}). This process is repeated until we obtain a solution where both $\tilde\chi(\tilde r)$ and $\tilde\varphi(\tilde r)$ vanish asymptotically. Using this numerical approach, we solve (\ref{eomHPbbres}) and present the results for two example cases, $d=5.5$ and $d=6.5$, in figure \ref{tchiphid55} and \ref{tchiphid65}, respectively.

From the numerical solutions, we find that for $d<6$, exemplified by the case $d=5.5$ shown in figure \ref{tchiphid55}, the amplitudes of $\tilde\chi$ and $\tilde\varphi$ increase with $m_\infty$. As $m_\infty\to 0$, both $\tilde\chi$ and $\tilde\varphi$ approach zero. In contrast, for $d>6$, as plotted in figure \ref{tchiphid65} for $d=6.5$, $\tilde\chi(1)$ and $\tilde\varphi(1)$ decrease as $m_\infty$ decreases, but they approach constants as $m_\infty\to 0$. This distinction in behavior between the cases $d<6$ and $d>6$ mirrors observations in flat space~\cite{Balthazar:2022hno}. Speaking of this, note that unlike the flat space case, Horowitz-Polchinski solutions in the thermal Kaluza-Kelin bubble can be found for $d>6$ without requiring the quartic terms in the effective action. Besides, in flat space, the solutions grow larger as $d-6$ increases, causing the effective action to break down~\cite{Balthazar:2022hno}. By contrast, as seen from (\ref{resHPbb}), in the thermal Kaluza-Klein bubble background, the solutions remain small as long as $r_H\gg l_s$, which is also the condition for the validity of the background solution (\ref{tbb}), as discussed earlier.

\section{Derivation of the $\lambda^4$ term in the perturbed free energy}\label{secla4}
In this appendix, we derive the $\lambda^4$ term in the free energy (\ref{Flambda4}) at $m_{\infty,\text{GL}}$. To do so, it is necessary to solve the second-order perturbations, which involves the four fields introduced in (\ref{del2}), namely $ \chi_0$, $\varphi_0$, $\chi_2$ and $\varphi_2$.

By plugging
\begin{equation}
    \chi(r,z)=\chi_*(r)+ \lambda\chi_{1,\text{GL}}(r)\cos k_z z+\delta\chi^{(2)}\ ,\ \varphi(r)=\varphi_*(r)+\lambda\varphi_{1,\text{GL}}\cos k_zz+ \delta\varphi^{(2)}
\end{equation}
in the action (\ref{Sphichiz}) and integrating over $z$, we obtain a $d$-dimensional action for these four fields. At leading order, $O(\lambda^4)$, $\chi_0$ and $\varphi_0$ decouple from $\chi_2$ and $\varphi_2$. Specifically, the action takes the form
\begin{equation}
\label{Id(2)}
    \delta I_d^{(2)}=\delta I_d (\chi_0,\varphi_0)+\delta I_d (\chi_2,\varphi_2)\ .
\end{equation}
More explicitly, we find
\begin{equation}
\label{Id0}
\begin{split}
    \delta I_d(\chi_0,\varphi_0)=&\frac{\lambda^4\beta }{16\pi G_N^{(d+1)}}\int d^dx\bigg[(\nabla \chi_0)^2+(\nabla \varphi_0)^2+(m_{\infty,\text{GL}}^2+\frac{\kappa}{\alpha'}\varphi_*)\chi_0^2\\
    &+\frac{2\kappa}{\alpha'}\chi_*\varphi_0\chi_0+\frac{\kappa}{\alpha'}\chi_{1,\text{GL}}\varphi_{1,\text{GL}}\chi_0+\frac{\kappa}{2\alpha'}\chi_{1,\text{GL}}^2\varphi_0\bigg]\ ,
\end{split}
\end{equation}
and
\begin{equation}
\label{Id2}
\begin{split}
    \delta I_d(\chi_2,\varphi_2)=&\frac{\lambda^4\beta}{32\pi G_N^{(d+1)}}\int d^dx\bigg[
   (\nabla\chi_2)^2+(\nabla\varphi_2)^2+(4k_z^2+m_{\infty,\text{GL}}^2+\frac{\kappa}{\alpha'}\varphi_*)\chi_2^2\\
   &+4k_z^2\varphi_2^2+\frac{2\kappa}{\alpha'}\chi_*\varphi_2\chi_2+\frac{\kappa}{\alpha'}\chi_{1,\text{GL}}\varphi_{1,\text{GL}}\chi_2+\frac{\kappa}{2\alpha'}\chi_{1,\text{GL}}^2\varphi_2\bigg]\ ,
\end{split}
\end{equation}
where we have neglected higher-order terms, such as $\varphi_0\chi_0^2$ which is of order $\lambda^6$. Note that (\ref{Id0}) and (\ref{Id2}) include source terms such as $\chi_{1,\text{GL}}\varphi_{1,\text{GL}}\chi_0$, which are linear in the second order perturbations.
\begin{figure}
	\centering
	\subfigure[]{
	\begin{minipage}[t]{0.45\linewidth}
	\centering
	\includegraphics[width=2.5in]{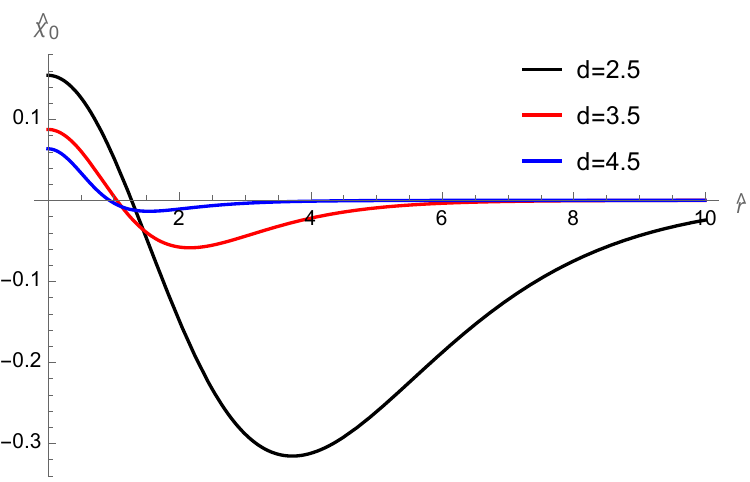}\label{delchi0}
	\end{minipage}}
	\subfigure[]{
	\begin{minipage}[t]{0.45\linewidth}
	\centering
	\includegraphics[width=2.5in]{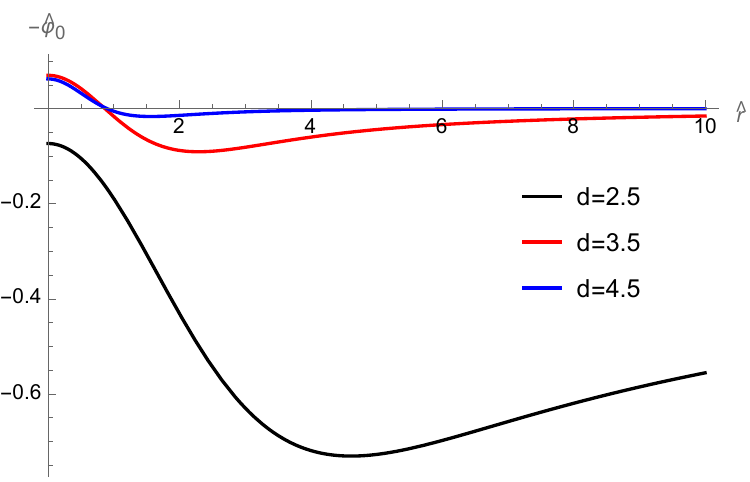}\label{delphi0}
	\end{minipage}}
	\centering
\caption{\label{dela}Profiles of the next-to-leading-order perturbations $\hat\chi_0$ and $-\hat\varphi_0$ for $d=2.5$, 3.5 and 4.5.}
\end{figure}
\begin{figure}
	\centering
	\subfigure[]{
	\begin{minipage}[t]{0.45\linewidth}
	\centering
	\includegraphics[width=2.5in]{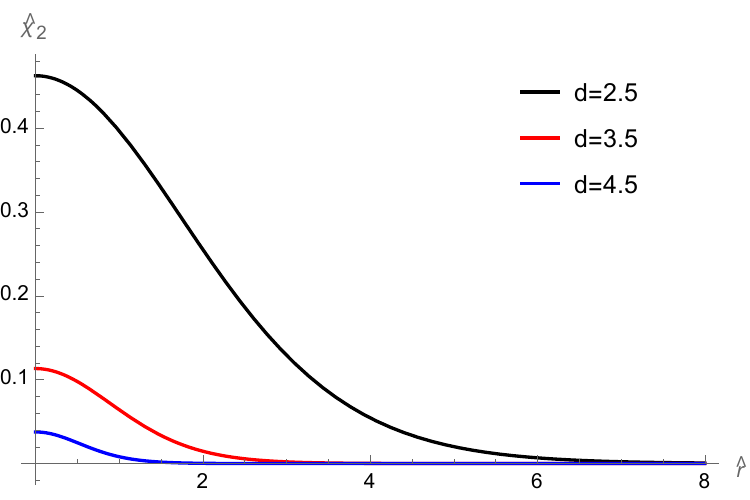}\label{delchi2}
	\end{minipage}}
	\subfigure[]{
	\begin{minipage}[t]{0.45\linewidth}
	\centering
	\includegraphics[width=2.5in]{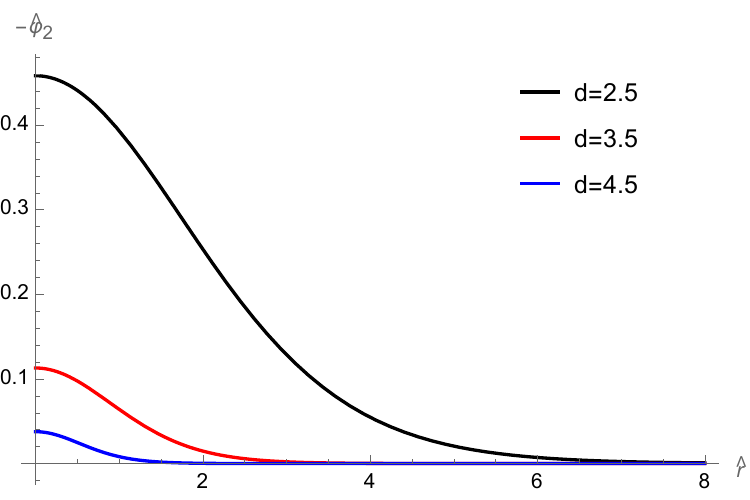}\label{delphi2}
	\end{minipage}}
	\centering
\caption{\label{delc}Profiles of the next-to-leading-order perturbations $\hat\chi_2$ and $-\hat\varphi_2$ for $d=2.5$, 3.5 and 4.5.}
\end{figure}

We solve the e.o.m. derived from (\ref{Id0}) and (\ref{Id2}) numerically, though for brevity we do not explicitly write them here. We impose the boundary conditions that the four fields $\chi_0$, $\varphi_0$, $\chi_2$ and $\varphi_2$ are finite at $r=0$ and vanish as $r\to \infty$. As before, it is useful to perform the rescaling (\ref{resHP}). The numerical solutions for $d=2.5$, 3.5 and 4.5 are plotted in figures \ref{dela} and \ref{delc}.

Using (\ref{Id0}) and (\ref{Id2}), we can now compute the $\lambda^4$ term in the free energy at $m_{\infty,\text{GL}}$. By integrating (\ref{Id(2)}) by parts and utilizing the equations of motion derived from it, we find that
\begin{equation}
\label{delI}
\begin{split}
    \delta F^{(2)}=\frac{\lambda^4\omega_{d-1}\kappa}{32\pi G_N^{(d+1)}\alpha'}\int dr\ r^{d-1}\bigg(&\chi_{1,\text{GL}}\varphi_{1,\text{GL}}\chi_0+\frac{1}{2}\chi_{1,\text{GL}}^2\varphi_0\\
    &+\frac{1}{2}\chi_{1,\text{GL}}\varphi_{1,\text{GL}}\chi_2+\frac{1}{4}\chi_{1,\text{GL}}^2\varphi_2\bigg)\\
    =\frac{\lambda^4\omega_{d-1}\alpha'^2}{16\pi G_N^{(d+1)}\kappa^2}m_{\infty,\text{GL}}^{6-d}\int d\hat r\ \hat r^{d-1}\bigg(&\hat\chi_{1,\text{GL}}\hat\varphi_{1,\text{GL}}\hat\chi_0+\frac{1}{2}\hat\chi_{1,\text{GL}}^2\hat\varphi_0\\
    &+\frac{1}{2}\hat\chi_{1,\text{GL}}\hat\varphi_{1,\text{GL}}\hat\chi_2+\frac{1}{4}\hat\chi_{1,\text{GL}}^2\hat\varphi_2\bigg)\ .    
\end{split}
\end{equation}
where in the last step we have performed the rescaling (\ref{resHP}).

\bibliographystyle{utphys}
\bibliography{paper}{}

\end{document}

%% file: sonerMacros.tex
\usepackage{amsmath,amssymb,amsfonts,mathtools,bm} 
\numberwithin{equation}{section}
\usepackage{physics} 
\usepackage[margin=10pt,font=small,labelfont=bf]{caption}
\usepackage[pdftex,dvipsnames]{xcolor}
\definecolor{darkblue}{rgb}{0.1,0.1,.7}
\usepackage{graphicx}  
\graphicspath{ {images/} }
\mathcode`l="8000
\begingroup
\makeatletter
\lccode`\~=`\l
\DeclareMathSymbol{\lsb@l}{\mathalpha}{letters}{`l}
\lowercase{\gdef~{\ifnum\the\mathgroup=\m@ne \ell \else \lsb@l \fi}}%
\endgroup
\def\<#1\>{\expval{#1}}

\newcommand{\undertl}[1]{\mathord{\vtop{\ialign{##\crcr
				$\hfil\displaystyle{#1}\hfil$\crcr\noalign{\kern1.5pt\nointerlineskip}
				$\hfil\tilde{}\hfil$\crcr\noalign{\kern1.5pt}}}}} 

\newcommand   \lptl{\raise .8ex\hbox{$^\leftarrow$} \hspace{-9pt} \partial} 
\newcommand   \lrptl{\raise .8ex\hbox{$^\leftrightarrow$} \hspace{-9pt} \partial} 
\def\be#1\ee{\begin{equation}\begin{aligned}#1\end{aligned}\end{equation}}
\makeatletter
\DeclareRobustCommand\bea{\@ifnextchar[{\@@bea}{\@bea}}
\def\@@bea[#1]#2\eea{\begin{subequations}\begin{align}#2\end{align}\label{#1}\end{subequations}}
\def\@bea#1\eea{\begin{subequations}\begin{align}#1\end{align}\end{subequations}}
\makeatother
 \catcode`,\active
 
 \catcode`\,12
 
\renewcommand \l  {\lambda}

%% file: black_string.bbl
\providecommand{\href}[2]{#2}\begingroup\raggedright\begin{thebibliography}{10}

\bibitem{Bowick:1985af}
M.~J. Bowick, L.~Smolin, and L.~C.~R. Wijewardhana, ``{Role of String
  Excitations in the Last Stages of Black Hole Evaporation},''
  \href{http://dx.doi.org/10.1103/PhysRevLett.56.424}{{\em Phys. Rev. Lett.}
  {\bfseries 56} (1986) 424}.

\bibitem{Susskind:1993ws}
L.~Susskind, ``{Some speculations about black hole entropy in string theory},''
  \href{http://arxiv.org/abs/hep-th/9309145}{{\ttfamily arXiv:hep-th/9309145}}.

\bibitem{Horowitz:1996nw}
G.~T. Horowitz and J.~Polchinski, ``{A Correspondence principle for black holes
  and strings},'' \href{http://dx.doi.org/10.1103/PhysRevD.55.6189}{{\em Phys.
  Rev. D} {\bfseries 55} (1997) 6189--6197},
  \href{http://arxiv.org/abs/hep-th/9612146}{{\ttfamily arXiv:hep-th/9612146}}.

\bibitem{Horowitz:1997jc}
G.~T. Horowitz and J.~Polchinski, ``{Selfgravitating fundamental strings},''
  \href{http://dx.doi.org/10.1103/PhysRevD.57.2557}{{\em Phys. Rev. D}
  {\bfseries 57} (1998) 2557--2563},
  \href{http://arxiv.org/abs/hep-th/9707170}{{\ttfamily arXiv:hep-th/9707170}}.

\bibitem{Sen:1995in}
A.~Sen, ``{Extremal black holes and elementary string states},''
  \href{http://dx.doi.org/10.1142/S0217732395002234}{{\em Mod. Phys. Lett. A}
  {\bfseries 10} (1995) 2081--2094},
  \href{http://arxiv.org/abs/hep-th/9504147}{{\ttfamily arXiv:hep-th/9504147}}.

\bibitem{Damour:1999aw}
T.~Damour and G.~Veneziano, ``{Selfgravitating fundamental strings and black
  holes},'' \href{http://dx.doi.org/10.1016/S0550-3213(99)00596-9}{{\em Nucl.
  Phys. B} {\bfseries 568} (2000) 93--119},
  \href{http://arxiv.org/abs/hep-th/9907030}{{\ttfamily arXiv:hep-th/9907030}}.

\bibitem{Khuri:1999ez}
R.~R. Khuri, ``{Selfgravitating strings and string / black hole
  correspondence},''
  \href{http://dx.doi.org/10.1016/S0370-2693(99)01265-4}{{\em Phys. Lett. B}
  {\bfseries 470} (1999) 73--76},
  \href{http://arxiv.org/abs/hep-th/9910122}{{\ttfamily arXiv:hep-th/9910122}}.

\bibitem{Kutasov:2005rr}
D.~Kutasov, ``{Accelerating branes and the string/black hole transition},''
  \href{http://arxiv.org/abs/hep-th/0509170}{{\ttfamily arXiv:hep-th/0509170}}.

\bibitem{Giveon:2005jv}
A.~Giveon and D.~Kutasov, ``{The Charged black hole/string transition},''
  \href{http://dx.doi.org/10.1088/1126-6708/2006/01/120}{{\em JHEP} {\bfseries
  01} (2006) 120}, \href{http://arxiv.org/abs/hep-th/0510211}{{\ttfamily
  arXiv:hep-th/0510211}}.

\bibitem{Giveon:2006pr}
A.~Giveon and D.~Kutasov, ``{Fundamental strings and black holes},''
  \href{http://dx.doi.org/10.1088/1126-6708/2007/01/071}{{\em JHEP} {\bfseries
  01} (2007) 071}, \href{http://arxiv.org/abs/hep-th/0611062}{{\ttfamily
  arXiv:hep-th/0611062}}.

\bibitem{Brustein:2021cza}
R.~Brustein and Y.~Zigdon, ``{Black hole entropy sourced by string winding
  condensate},'' \href{http://dx.doi.org/10.1007/JHEP10(2021)219}{{\em JHEP}
  {\bfseries 10} (2021) 219}, \href{http://arxiv.org/abs/2107.09001}{{\ttfamily
  arXiv:2107.09001 [hep-th]}}.

\bibitem{Chen:2021emg}
Y.~Chen and J.~Maldacena, ``{String scale black holes at large D},''
  \href{http://dx.doi.org/10.1007/JHEP01(2022)095}{{\em JHEP} {\bfseries 01}
  (2022) 095}, \href{http://arxiv.org/abs/2106.02169}{{\ttfamily
  arXiv:2106.02169 [hep-th]}}.

\bibitem{Chen:2021dsw}
Y.~Chen, J.~Maldacena, and E.~Witten, ``{On the black hole/string
  transition},'' \href{http://dx.doi.org/10.1007/JHEP01(2023)103}{{\em JHEP}
  {\bfseries 01} (2023) 103}, \href{http://arxiv.org/abs/2109.08563}{{\ttfamily
  arXiv:2109.08563 [hep-th]}}.

\bibitem{Urbach:2022xzw}
E.~Y. Urbach, ``{String stars in anti de Sitter space},''
  \href{http://dx.doi.org/10.1007/JHEP04(2022)072}{{\em JHEP} {\bfseries 04}
  (2022) 072}, \href{http://arxiv.org/abs/2202.06966}{{\ttfamily
  arXiv:2202.06966 [hep-th]}}.

\bibitem{Balthazar:2022szl}
B.~Balthazar, J.~Chu, and D.~Kutasov, ``{Winding Tachyons and Stringy Black
  Holes},'' \href{http://arxiv.org/abs/2204.00012}{{\ttfamily arXiv:2204.00012
  [hep-th]}}.

\bibitem{Balthazar:2022hno}
B.~Balthazar, J.~Chu, and D.~Kutasov, ``{On small black holes in string
  theory},'' \href{http://dx.doi.org/10.1007/JHEP03(2024)116}{{\em JHEP}
  {\bfseries 03} (2024) 116}, \href{http://arxiv.org/abs/2210.12033}{{\ttfamily
  arXiv:2210.12033 [hep-th]}}.

\bibitem{Bedroya:2022twb}
A.~Bedroya, ``{High energy scattering and string/black hole transition},''
  \href{http://arxiv.org/abs/2211.17162}{{\ttfamily arXiv:2211.17162
  [hep-th]}}.

\bibitem{Urbach:2023npi}
E.~Y. Urbach, ``{The black hole/string transition in AdS$_{3}$ and confining
  backgrounds},'' \href{http://dx.doi.org/10.1007/JHEP09(2023)156}{{\em JHEP}
  {\bfseries 09} (2023) 156}, \href{http://arxiv.org/abs/2303.09567}{{\ttfamily
  arXiv:2303.09567 [hep-th]}}.

\bibitem{Ceplak:2023afb}
N.~\v{C}eplak, R.~Emparan, A.~Puhm, and M.~Toma\v{s}evi\'c, ``{The
  correspondence between rotating black holes and fundamental strings},''
  \href{http://dx.doi.org/10.1007/JHEP11(2023)226}{{\em JHEP} {\bfseries 11}
  (2023) 226}, \href{http://arxiv.org/abs/2307.03573}{{\ttfamily
  arXiv:2307.03573 [hep-th]}}.

\bibitem{Halder:2023nlp}
I.~Halder and D.~L. Jafferis, ``{Double winding condensate CFT},''
  \href{http://dx.doi.org/10.1007/JHEP05(2024)189}{{\em JHEP} {\bfseries 05}
  (2024) 189}, \href{http://arxiv.org/abs/2308.11702}{{\ttfamily
  arXiv:2308.11702 [hep-th]}}.

\bibitem{Agia:2023skp}
N.~Agia and D.~L. Jafferis, ``{AdS$_3$ String Stars at Pure NSNS Flux},''
  \href{http://arxiv.org/abs/2311.04956}{{\ttfamily arXiv:2311.04956
  [hep-th]}}.

\bibitem{Bedroya:2024uva}
A.~Bedroya, C.~Vafa, and D.~H. Wu, ``{The Tale of Three Scales: the Planck, the
  Species, and the Black Hole Scales},''
  \href{http://arxiv.org/abs/2403.18005}{{\ttfamily arXiv:2403.18005
  [hep-th]}}.

\bibitem{Santos:2024ycg}
J.~E. Santos and Y.~Zigdon, ``{Self gravitating spinning string condensates},''
  \href{http://dx.doi.org/10.1007/JHEP07(2024)217}{{\em JHEP} {\bfseries 07}
  (2024) 217}, \href{http://arxiv.org/abs/2403.20332}{{\ttfamily
  arXiv:2403.20332 [hep-th]}}.

\bibitem{Gregory:1993vy}
R.~Gregory and R.~Laflamme, ``{Black strings and p-branes are unstable},''
  \href{http://dx.doi.org/10.1103/PhysRevLett.70.2837}{{\em Phys. Rev. Lett.}
  {\bfseries 70} (1993) 2837--2840},
  \href{http://arxiv.org/abs/hep-th/9301052}{{\ttfamily arXiv:hep-th/9301052}}.

\bibitem{Gregory:1994bj}
R.~Gregory and R.~Laflamme, ``{The Instability of charged black strings and
  p-branes},'' \href{http://dx.doi.org/10.1016/0550-3213(94)90206-2}{{\em Nucl.
  Phys. B} {\bfseries 428} (1994) 399--434},
  \href{http://arxiv.org/abs/hep-th/9404071}{{\ttfamily arXiv:hep-th/9404071}}.

\bibitem{Kol:2004ww}
B.~Kol, ``{The Phase transition between caged black holes and black strings: A
  Review},'' \href{http://dx.doi.org/10.1016/j.physrep.2005.10.001}{{\em Phys.
  Rept.} {\bfseries 422} (2006) 119--165},
  \href{http://arxiv.org/abs/hep-th/0411240}{{\ttfamily arXiv:hep-th/0411240}}.

\bibitem{Emparan:2013xia}
R.~Emparan, D.~Grumiller, and K.~Tanabe, ``{Large-D gravity and low-D
  strings},'' \href{http://dx.doi.org/10.1103/PhysRevLett.110.251102}{{\em
  Phys. Rev. Lett.} {\bfseries 110} no.~25, (2013) 251102},
  \href{http://arxiv.org/abs/1303.1995}{{\ttfamily arXiv:1303.1995 [hep-th]}}.

\bibitem{Halder:2024gwe}
I.~Halder and D.~L. Jafferis, ``{Stretched horizon, replica trick and off-shell
  winding condensate, and all that},''
  \href{http://arxiv.org/abs/2402.00932}{{\ttfamily arXiv:2402.00932
  [hep-th]}}.

\bibitem{CALLAN1989673}
C.~Callan, R.~Myers, and M.~Perry, ``Black holes in string theory,''
  \href{http://dx.doi.org/https://doi.org/10.1016/0550-3213(89)90172-7}{{\em
  Nuclear Physics B} {\bfseries 311} no.~3, (1989) 673--698}.
  \url{https://www.sciencedirect.com/science/article/pii/0550321389901727}.

\bibitem{Polchinski:1985zf}
J.~Polchinski, ``{Evaluation of the One Loop String Path Integral},''
  \href{http://dx.doi.org/10.1007/BF01210791}{{\em Commun. Math. Phys.}
  {\bfseries 104} (1986) 37}.

\bibitem{Atick:1988si}
J.~J. Atick and E.~Witten, ``{The Hagedorn Transition and the Number of Degrees
  of Freedom of String Theory},''
  \href{http://dx.doi.org/10.1016/0550-3213(88)90151-4}{{\em Nucl. Phys. B}
  {\bfseries 310} (1988) 291--334}.

\bibitem{Brustein:2021ifl}
R.~Brustein and Y.~Zigdon, ``{Effective field theory for closed strings near
  the Hagedorn temperature},''
  \href{http://dx.doi.org/10.1007/JHEP04(2021)107}{{\em JHEP} {\bfseries 04}
  (2021) 107}, \href{http://arxiv.org/abs/2101.07836}{{\ttfamily
  arXiv:2101.07836 [hep-th]}}.

\bibitem{Emparan:2018bmi}
R.~Emparan, R.~Luna, M.~Mart\'\i{}nez, R.~Suzuki, and K.~Tanabe, ``{Phases and
  Stability of Non-Uniform Black Strings},''
  \href{http://dx.doi.org/10.1007/JHEP05(2018)104}{{\em JHEP} {\bfseries 05}
  (2018) 104}, \href{http://arxiv.org/abs/1802.08191}{{\ttfamily
  arXiv:1802.08191 [hep-th]}}.

\bibitem{Sorkin:2004qq}
E.~Sorkin, ``{A Critical dimension in the black string phase transition},''
  \href{http://dx.doi.org/10.1103/PhysRevLett.93.031601}{{\em Phys. Rev. Lett.}
  {\bfseries 93} (2004) 031601},
  \href{http://arxiv.org/abs/hep-th/0402216}{{\ttfamily arXiv:hep-th/0402216}}.

\bibitem{Gubser:2000ec}
S.~S. Gubser and I.~Mitra, ``{Instability of charged black holes in Anti-de
  Sitter space},'' {\em Clay Math. Proc.} {\bfseries 1} (2002) 221,
  \href{http://arxiv.org/abs/hep-th/0009126}{{\ttfamily arXiv:hep-th/0009126}}.

\bibitem{Gubser:2000mm}
S.~S. Gubser and I.~Mitra, ``{The Evolution of unstable black holes in anti-de
  Sitter space},'' \href{http://dx.doi.org/10.1088/1126-6708/2001/08/018}{{\em
  JHEP} {\bfseries 08} (2001) 018},
  \href{http://arxiv.org/abs/hep-th/0011127}{{\ttfamily arXiv:hep-th/0011127}}.

\bibitem{Bostock:2004mg}
P.~Bostock and S.~F. Ross, ``{Smeared branes and the Gubser-Mitra
  conjecture},'' \href{http://dx.doi.org/10.1103/PhysRevD.70.064014}{{\em Phys.
  Rev. D} {\bfseries 70} (2004) 064014},
  \href{http://arxiv.org/abs/hep-th/0405026}{{\ttfamily arXiv:hep-th/0405026}}.

\bibitem{Marolf:2004fya}
D.~Marolf and B.~Cabrera~Palmer, ``{Gyrating strings: A New instability of
  black strings?},'' \href{http://dx.doi.org/10.1103/PhysRevD.70.084045}{{\em
  Phys. Rev. D} {\bfseries 70} (2004) 084045},
  \href{http://arxiv.org/abs/hep-th/0404139}{{\ttfamily arXiv:hep-th/0404139}}.

\bibitem{Friess:2005zp}
J.~J. Friess, S.~S. Gubser, and I.~Mitra, ``{Counter-examples to the correlated
  stability conjecture},''
  \href{http://dx.doi.org/10.1103/PhysRevD.72.104019}{{\em Phys. Rev. D}
  {\bfseries 72} (2005) 104019},
  \href{http://arxiv.org/abs/hep-th/0508220}{{\ttfamily arXiv:hep-th/0508220}}.

\bibitem{Giveon:2005mi}
A.~Giveon, D.~Kutasov, E.~Rabinovici, and A.~Sever, ``{Phases of quantum
  gravity in AdS(3) and linear dilaton backgrounds},''
  \href{http://dx.doi.org/10.1016/j.nuclphysb.2005.04.015}{{\em Nucl. Phys. B}
  {\bfseries 719} (2005) 3--34},
  \href{http://arxiv.org/abs/hep-th/0503121}{{\ttfamily arXiv:hep-th/0503121}}.

\bibitem{Balthazar:2021xeh}
B.~Balthazar, A.~Giveon, D.~Kutasov, and E.~J. Martinec, ``{Asymptotically free
  AdS$_{3}$/CFT$_{2}$},'' \href{http://dx.doi.org/10.1007/JHEP01(2022)008}{{\em
  JHEP} {\bfseries 01} (2022) 008},
  \href{http://arxiv.org/abs/2109.00065}{{\ttfamily arXiv:2109.00065
  [hep-th]}}.

\bibitem{Asnin:2007rw}
V.~Asnin, D.~Gorbonos, S.~Hadar, B.~Kol, M.~Levi, and U.~Miyamoto, ``{High and
  Low Dimensions in The Black Hole Negative Mode},''
  \href{http://dx.doi.org/10.1088/0264-9381/24/22/015}{{\em Class. Quant.
  Grav.} {\bfseries 24} (2007) 5527--5540},
  \href{http://arxiv.org/abs/0706.1555}{{\ttfamily arXiv:0706.1555 [hep-th]}}.

\bibitem{Kol:2004pn}
B.~Kol and E.~Sorkin, ``{On black-brane instability in an arbitrary
  dimension},'' \href{http://dx.doi.org/10.1088/0264-9381/21/21/003}{{\em
  Class. Quant. Grav.} {\bfseries 21} (2004) 4793--4804},
  \href{http://arxiv.org/abs/gr-qc/0407058}{{\ttfamily arXiv:gr-qc/0407058}}.

\bibitem{Emparan:2024mbp}
R.~Emparan, M.~Sanchez-Garitaonandia, and M.~Toma\v{s}evi\'c, ``{String Theory
  in a Pinch: Resolving the Gregory-Laflamme Singularity},''
  \href{http://arxiv.org/abs/2411.14998}{{\ttfamily arXiv:2411.14998
  [hep-th]}}.

\bibitem{Elvang:2004iz}
H.~Elvang, T.~Harmark, and N.~A. Obers, ``{Sequences of bubbles and holes: New
  phases of Kaluza-Klein black holes},''
  \href{http://dx.doi.org/10.1088/1126-6708/2005/01/003}{{\em JHEP} {\bfseries
  01} (2005) 003}, \href{http://arxiv.org/abs/hep-th/0407050}{{\ttfamily
  arXiv:hep-th/0407050}}.

\bibitem{Kol:2006vu}
B.~Kol and E.~Sorkin, ``{LG (Landau-Ginzburg) in GL (Gregory-Laflamme)},''
  \href{http://dx.doi.org/10.1088/0264-9381/23/14/002}{{\em Class. Quant.
  Grav.} {\bfseries 23} (2006) 4563--4592},
  \href{http://arxiv.org/abs/hep-th/0604015}{{\ttfamily arXiv:hep-th/0604015}}.

\bibitem{Aharony:2004ig}
O.~Aharony, J.~Marsano, S.~Minwalla, and T.~Wiseman, ``{Black hole-black string
  phase transitions in thermal 1+1 dimensional supersymmetric Yang-Mills theory
  on a circle},'' \href{http://dx.doi.org/10.1088/0264-9381/21/22/010}{{\em
  Class. Quant. Grav.} {\bfseries 21} (2004) 5169--5192},
  \href{http://arxiv.org/abs/hep-th/0406210}{{\ttfamily arXiv:hep-th/0406210}}.

\bibitem{Harmark:2004ws}
T.~Harmark and N.~A. Obers, ``{New phases of near-extremal branes on a
  circle},'' \href{http://dx.doi.org/10.1088/1126-6708/2004/09/022}{{\em JHEP}
  {\bfseries 09} (2004) 022},
  \href{http://arxiv.org/abs/hep-th/0407094}{{\ttfamily arXiv:hep-th/0407094}}.

\bibitem{Witten:1981gj}
E.~Witten, ``{Instability of the Kaluza-Klein Vacuum},''
  \href{http://dx.doi.org/10.1016/0550-3213(82)90007-4}{{\em Nucl. Phys. B}
  {\bfseries 195} (1982) 481--492}.

\end{thebibliography}\endgroup
